\documentclass[twocolumn]{aa}
\usepackage{graphicx}
\usepackage{natbib}

\usepackage{txfonts}
\begin{document}
    \title{X-ray Emission from T Tauri Stars and the Role of Accretion: Inferences 
           from the XMM-Newton Extended Survey of the Taurus Molecular Cloud}
    \titlerunning{X-ray emission from T Tau stars and the role of accretion}

   \author{Alessandra Telleschi\inst{1}, Manuel G\"udel\inst{1}, Kevin~R. Briggs\inst{1},
          Marc Audard\inst{2} \and Francesco Palla\inst{3}}
   \authorrunning{A. Telleschi et al.}
   \offprints{A. Telleschi}

   \institute{Paul Scherrer Institut, W\"urenlingen and Villigen,
              CH-5232 Villigen PSI, Switzerland
              \and
             Columbia Astrophysics Laboratory, Columbia University, 550 West 120th Street, Mail code 5247, New York, NY 10027, USA
             \fnmsep\thanks{\emph{New address (since September 2006): }Integral Science Data Centre, Ch. d'Ecogia 16, CH-1290 Versoix, Switzerland \& Geneva Observatory, University of Geneva, Ch. des Maillettes 5
1, 1290 Sauverny, Switzerland}
             \and
             INAF-Osservatorio Astrofisico di Arcetri,
             Largo Enrico Fermi, 5,
             I-50125 Firenze, Italy}

   \date{Recived 2006; accepted 2006}

  \abstract
   {T Tau stars display different X-ray properties
   depending on whether they are accreting (classical T Tau stars; CTTS) or not (weak-line T Tau stars; WTTS).
   X-ray properties may provide insight into the accretion process between disk and stellar surface.}
   {We use data from the {{\it XMM-Newton Extended Survey of the Taurus Molecular Cloud}} (XEST) to 
   study differences in X-ray properties between CTTS and WTTS.}
   {XEST data are used to perform correlation and regression analysis between X-ray parameters and
   stellar properties.}
   {We confirm the existence of a X-ray luminosity ($L_{\rm X}$) vs. mass $(M)$ relation,
   $L_{\rm X}\propto M^{1.69\pm 0.11}$, but this relation is a consequence
   of X-ray saturation and a mass vs. bolometric luminosity ($L_*$) relation for the TTS
   with an average age of 2.4~Myr. X-ray saturation 
   indicates $ L_{\rm X} = {\rm const}L_*$, although the constant is different for the two
   subsamples: const = $10^{-3.73\pm 0.05}$ for CTTS and const = $10^{-3.39\pm 0.06}$ for WTTS. Given a 
   similar $L_*$ distribution of both samples, the X-ray luminosity function also reflects a real
   X-ray deficiency in CTTS, by a factor of $\approx 2$ compared to WTTS. 
   The average electron temperatures $T_{\rm av}$ are correlated with $L_{\rm X}$ in WTTS but not in CTTS;
   CTTS sources are on average hotter than WTTS sources. At best marginal dependencies are found
   between X-ray properties and mass accretion rates or age.}
   {The most fundamental properties are the two saturation laws,
   indicating suppressed $L_{\rm X}$  for CTTS. We speculate that some of the accreting material
   in CTTS is cooling active regions to temperatures that may not significantly emit in the X-ray band, and if 
   they do, high-resolution spectroscopy may be required to identify lines formed in such plasma, while 
   CCD cameras do not detect these components. The similarity of the $L_{\rm X}$ vs. $T_{\rm av}$ dependencies
   in WTTS and main-sequence stars as well as their similar X-ray saturation laws suggests similar physical 
   processes for the hot plasma, i.e., heating and radiation of a magnetic corona.   }


   \maketitle
%
\section{Introduction}

Optically revealed low mass pre-main-sequence stars define the class 
of T Tauri Stars (TTS). TTS are divided into two families, the Classical 
T Tauri Stars (CTTS) and the Weak-line T Tauri Stars (WTTS). CTTS 
display strong H$\alpha$ lines, a sign that the stars are accreting
material from the circumstellar disk, while in WTTS the H$\alpha$
line fluxes are suppressed, a sign that accretion has ceased. Based on 
infrared observations, Young Stellar Objects (YSO) have instead been  
ordered in classes according to their infrared (IR) excess. Following this
classification, deeply embedded stars at the start of their accretion 
phase are ``Class 0'' objects, more evolved protostars still embedded in 
their envelope are ``Class I'' objects, stars with a circumstellar disk
that show IR excess are ``Class II'' objects, and stars with no IR
excess are ``Class III'' objects. While the H$\alpha$
classification is based on accretion, the IR excess is a measure of
circumstellar material. The Class II objects are dominated by CTTS, 
while the Class III stars are dominated by WTTS. 

Both types of TTS have been  found to be strong X-ray emitters. 
First X-ray detections of individual TTS were made with the {\it Einstein}  
observatory \citep[e.g.,][]{feigelson81} and revealed very
strong X-ray activity, exceeding the solar level by several orders of magnitude.
Many star-forming regions have subsequently been observed with the 
{\it ROSAT} satellite \citep[e.g.,][]{feigelson93,gagne95,
neuhaeuser95,stelzer01}, largely increasing the number of X-ray detected
TTS. Studies based on H$\alpha$ emission may in fact fail to detect part of 
the WTTS population, which can  easily be identified in X-rays.

The origin of the strong X-ray activity in TTS is not entirely clear.
The observed emission in the  soft X-ray band above 1~keV
is consistent with emission from a scaled-up 
version of the solar corona.
In main-sequence stars X-ray activity is mainly determined
by the stellar rotation rate. The activity-rotation relation is given
by $L_{\rm X}/L_* \propto P_{\rm rot}^{-2.6}$ \citep{guedel97},
where $L_{\rm X}$ is the X-ray luminosity, $L_*$ is the stellar photospheric
bolometric luminosity, and $P_{\rm rot}$ is the rotation period of
the star.
This is consistent with the dynamo mechanism that is present in our Sun,
where the magnetic fields are generated through an $\alpha$-$\Omega$
dynamo \citep{parker55}. 
At rotation periods shorter than 2-3 days for G-K stars, the X-ray activity 
saturates at $\log(L_{\rm X}/L_*) \approx -3$ \citep{vilhu83}.

As for pre-main sequence stars, early surveys of the Taurus Molecular Cloud (TMC;
\citealt{neuhaeuser95,stelzer01}) claimed a rotation-activity
relation somewhat similar to the relation for main-sequence stars, but
the recent COUP survey of the Orion Nebula Cluster (ONC) found absence of 
such a relation  \citep{preibisch05}, suggesting that all stars are in
a saturation regime, even for long rotation periods. Young stellar 
objects, especially  in their early  evolutionary stage, are thought to be fully convective,
and the generation of magnetic fields through the $\alpha$-$\Omega$
dynamo should not be possible.
This suggests that X-rays in low-mass pre-main sequence stars 
are generated through processes different than in the Sun. 
New models for X-ray generation through other dynamo 
concepts have been developed \citep{kueker99, giampapa96}.
Alternatively, in CTTS, X-rays could in principle be produced by
magnetic star-disk interactions \citep[e.g.,][]{montmerle00,
isobe03}, in accretion shocks \citep[e.g.,][]{lamzin99,kastner02, stelzer04},
or in shocks at the base of outflows and jets \citep{guedel05,kastner05}.

The influence of a circumstellar disk, and particularly the
influence of accretion on X-ray activity is therefore of interest. Former
X-ray studies of star forming regions have led to discrepant
results. In the Taurus-Auriga complex, \citet{stelzer01} reported
higher X-ray luminosities for the non-accreting WTTS stars than for CTTS.
In the ONC, \citet{feigelson02}  concluded from {\it Chandra} observations 
that the presence of circumstellar disks has no influence
on the X-ray emission, whereas \citet{flaccomio03a}, in another {\it Chandra}
study of the ONC, found $L_{\rm X}$ and
$L_{\rm X}/L_*$ to be enhanced in WTTS when compared to CTTS. From the recent
{\it Chandra} Orion Ultradeep Project (COUP), \citet{preibisch05}
reported the X-ray emission of WTTS to be
consistent with the X-ray emission of active Main Sequence (MS) 
stars, while it is suppressed in CTTS.
However, in all these studies the X-ray emission mechanism is
consistent with a scaled-up version of a solar corona.

X-ray emission during accretion outbursts has been
observed in V1647 Ori \citep{kastner04, grosso05, kastner06}
and in V1118 Ori \citep{audard05}. The X-ray luminosity 
increased by a factor of 50 during the outburst in V1647 Ori,
and the spectrum hardened. On the other hand, the X-ray
luminosity of V1118 Ori remained at the same level as during the
pre-outburst phase, while the spectrum became softer.

Possible signs of accretion-induced X-ray emission are revealed 
in a few high-resolution spectra of CTTS. High electron densities 
were measured in the spectra of TW Hya \citep{kastner02, stelzer04},
BP Tau \citep{schmitt05,robrade06}, and V4046 Sgr \citep{guenther06}, and were interpreted
as indications of X-ray production in accretion shocks. Other
spectroscopic features that also suggested an accretion shock
scenario are the low electron temperature dominating the plasma in TW Hya (a few MK,
as expected from shock-induced heating) and abundance anomalies. 
\citet{stelzer04} interpreted the high Ne/Fe abundance ratio 
as being due to Fe depletion by condensation into grain
in the accretion disk. \citet{drake05} reported
a substantially larger Ne/O ratio in the spectrum of TW Hya than
in the spectra of the other studied stars, and they proposed to
use this ratio as a diagnostic for metal depletion in the
circumstellar disk of accreting stars.

Work on high-resolution X-ray spectroscopy was subsequently extended by
\citet{telleschi06a} to a sample of 9 pre-main sequence stars with different accretion
properties. The main result of that work is the identification of an
excess of cool plasma measured in the accreting stars, when compared
to WTTS. The origin of this soft excess is unclear.   
Further evidence for a strong soft excess in the CTTS is revealed in the extraordinary 
X-ray spectrum of T Tau \citep{guedel06c}. In this case, however, the electron 
density (derived from spectral lines formed at low temperatures) is 
low, $ n_{\rm e} \lesssim 10^{10}$~cm$^{-3}$. The density, in case of
accretion shocks, can be estimated using the strong shock condition $n_2=4n_1$, 
where $n_1$ and $n_2$ are the pre-shock and post-shock densities, respectively.
The density $n_1$ can be derived from the accretion mass rate and the accreting
area on the stellar surface: $\dot{M} \approx 4\pi R^2fv_{ff}n_em_p$,
where $f$ is the surface filling factor of the accretion flow, and 
$v_{ff}=(2GM/R)^{1/2}$ is the free-fall velocity. Using the accretion
rate $\dot{M} \approx (3-6) \times 10^{-8}~M_{\odot}$~yr$^{-1}$ for T Tau 
\citep{white01, calvet04} we obtain $n_2 = (1.1 -2.2) \times 10^{11}/f$
\citep{guedel06c}. Even in the extreme case that $f = 10$\%, we
expect a density $\gtrsim 10^{12}$~cm$^{-3}$, i.e. orders of magnitude 
higher than the measured value.


The aim of the present paper is to study the role of accretion
in the overall X-ray properties of  pre-main sequence stars in the 
Taurus-Auriga Molecular Cloud, by coherently comparing samples of  CTTS 
and WTTS.  Our analysis  is complementary to the
COUP survey work, and we will present our results
along largely similar lines (see \citealt{preibisch05} for COUP). Indeed 
one of the main purposes of the present work is  a 
qualitative comparison of the Taurus results with those
obtained from the Orion sample. We do not,
however, present issues related to rotation; rotation-activity
relations will be separately discussed in a dedicated paper 
\citep{briggs06}.
 
The paper is structured as follows. In Sect.~\ref{sample} we describe 
the stellar sample used in this work, and in \ref{data} we summarize the
relevant steps of the data reduction. We present our results in Sect.~\ref{res},
and discuss them in Sect.~\ref{discussion}. We summarize our results
and conclude in Sect.~\ref{conclusions}.

\section{Studying X-rays in the Taurus Molecular Cloud}\label{sample}

\subsection{The Taurus Molecular Cloud}

We will address questions on X-ray production in accreting and non-accreting
T Tauri stars using data from the {\it XMM-Newton Extended Survey of the
Taurus Molecular Cloud} (XEST, \citealt{guedel06a}).

The Taurus Molecular Cloud (TMC) varies in significant ways from the Orion 
Nebula Cluster and makes our study an important complement to the COUP
survey. The TMC has, as the nearest large, star-forming region (distance 
$\approx$ 140~pc, \citealt{loinard05}), played a fundamental role in our 
understanding of low-mass star formation. It features several loosely
associated but otherwise rather isolated molecular cores, each of which 
produces one or only a few low-mass stars, different from the much denser 
cores in $\rho$ Oph or in Orion. TMC shows a low  stellar density of only 
1--10 stars~pc$^{-2}$ (e.g., \citealt{gomez93}). In contrast 
to the very dense environment in the Orion Nebula Cluster, strong mutual 
influence due to outflows, jets, or gravitational effects is therefore 
minimized. Also, strong winds and UV radiation fields of
OB stars are present in Orion but absent in the TMC.

The TMC has provided the best-characterized sample of CTTS and WTTS, many of
which have been subject to detailed studies; see, e.g., the seminal
work by \citet{kenyon95} that concerns, among other things, the
evolutionary history of T Tau stars and their disk+envelope environment, 
mostly based on optical and infrared observations.

It is therefore little surprising that comprehensive X-ray studies of 
selected objects as well as surveys have been performed with several previous  
X-ray satellites; for X-ray survey work see, e.g., the papers by  
\citet{feigelson87}, \citet{walter88}, \citet{bouvier90},  \citet{strom90}, 
\citet{strom94}, \citet{damiani95a}, \citet{damiani95b},
\citet{neuhaeuser95}, and  \citet{stelzer01}. Issues we are studying
in our paper have variously been studied in these surveys before, although, as
argued by \citet{guedel06a} and below, the present XEST project
is more sensitive and provides us with a near-complete sample of X-ray detected
TTS in the surveyed area, thus minimizing selection and detection bias.

\subsection{The XEST sample of T Tau stars}

The XEST project is an X-ray study of the most 
populated regions (comprising an area of $\approx$ 5 square degrees) of the Taurus 
Molecular Cloud. The survey consists of 28 {\it XMM-Newton}
exposures. The 19 initial observations of the  
project (of approximately 30 ks duration each, see Table~1 of \citealt{guedel06a}) 
were complemented by 9 exposures from other projects
or from the archive.
Also, 6 {\it Chandra} observations have been used in XEST, 
to add information on a few sources not detected 
with {\it XMM-Newton}, or binary information (see \citealt{guedel06a}).

To distinguish between CTTS and WTTS we use the classification
given in col.~10 of Table~11 of \citet{guedel06a}. This 
classification is substantially based on the equivalent
width of the H$\alpha$ line (EW[H$\alpha$]). For spectral types G and K,
stars with EW(H$\alpha$) $\ge$ 5 \AA\ are defined as CTTS,
while other stars are defined as WTTS. For early-M spectral types,
the boundary between CTTS and WTTS was set at EW(H$\alpha$) = 10 \AA\
and for mid-M spectral types  at EW(H$\alpha$) = 20 \AA.
Stars with late-M spectral type are mainly Brown Dwarfs (BDs),
and given their low optical continuum, a clear accretion criterion is 
difficult to provide. 
For this reason, BDs were treated as a class of their
own and are not used in our comparison studies of
accretors vs. non-accretors (but were included  in the ``total'' samples when appropriate). 
For further details, see \citet{guedel06a} and \citet{grosso06}.
YSO IR types were used to classify borderline cases and protostars.
In summary, protostars have been classified as type 0 or 1 (Class 0 and I, respectively),
CTTS are type 2 objects, WTTS are type 3 objects,
and BDs are classified as type 4. Type 5 is assigned to Herbig Ae/Be stars, while stars
with uncertain classification are assigned to type 9. We will use these
designations in our illustrations below.

We emphasize the near-completeness of XEST with regard to X-ray detections
of TTS. \citet{guedel06a} provide the detection statistics (their Table 12): 
A total of 126 out of the 159 TMC members surveyed with {\it XMM-Newton} were 
detected in X-rays. Among these are  55 detected CTTS and 49 detected WTTS  (out of
the 65 and 50 surveyed targets), corresponding to a detection fraction of 
85\% and 98\%, respectively. Almost all objects
have been found comfortably above the approximate detection limit of 
$L_{\rm X} \approx 10^{28}$~erg~s$^{-1}$, indicating that TTS generally
emit at levels  between $10^{29} - 10^{31}$~erg~s$^{-1}$, exceptions
being lowest-mass stars and brown dwarfs. Most of the non-detected objects
have been recognized as stars that are strongly absorbed (e.g., by their 
own disks) or as stars of very low mass \citep{guedel06a}. XEST is the first X-ray survey
of TMC that reaches completeness fractions near unity, and therefore minimizes 
detection bias and unknown effects of upper limits to correlation studies as 
performed here. It provides,  in this regard, an ideal comparison with the COUP
results \citep{preibisch05}.

A few sources were excluded from consideration in
the present work. These are the four stars that show composite 
X-ray spectra possibly originating from two different sources (DG Tau A, 
GV Tau, DP Tau, and CW Tau; \citealt{guedel06b}), and three stars which show a 
decreasing light curve throughout the observation (DH Tau, FS Tau AC, and V830 Tau); these light
curves probably describe the late phases of  large  flares. 
Further, the deeply embedded protostar L1551 IRS5, which shows 
lightly absorbed X-ray emission that may be attributed to the 
jets \citep{favata02,bally03}, was also excluded, and so were the two 
Herbig stars (AB Aur, V892 Tau). 
In some correlation studies, we do consider objects
for which upper limits to $L_{\rm X}$ have been estimated in \citet{guedel06a}, but will
not consider non-detections without such estimate (as, for example, if the absorption 
is unknown).

Our final, basic sample of TTS then consists of 56 CTTS and 49 WTTS. Among the X-ray
detections, there are also 8 protostars, 8 BDs, 2 Herbig stars and 4 stars with 
uncertain classification. Smaller subsamples may be used if
parameters of interest were not available.  

When $L_*$ is involved in a correlation, 
we excluded all stars that are apparently located below 
the Zero-Age Main Sequence (ZAMS) in the Hertzsprung-Russel diagram 
(Fig.~10 of \citealt{guedel06a}).
These stars are HH 30, IRAS S04301+261, Haro 6-5 B, HBC 353, and HBC 352.  
Their location in the HRD is likely to be due to inaccurate photometry.

Many of the stellar counterparts to our X-ray sources are unresolved binaries or multiples. In total,
45 out of the 159 stellar systems surveyed by {\it XMM-Newton} are multiple
\citep{guedel06a}. If - as we will find below in general, and as has
been reported in earlier studies of T Tauri stars \citep[e.g.,][]{preibisch05} -
$L_{\rm X}$ scales with $L_*$, then this also holds for the sum of
the $L_{\rm X}$ with respect to the sum of the component $L_*$.
Binarity does therefore not influence comparisons between $L_{\rm X}$
and $L_*$. When correlating $L_{\rm X}$ with stellar mass, we will
find that more massive stars are in general brighter. In the case of binaries,
the more massive component (usually the more luminous ``primary'' star) will
thus dominate the X-ray emission. We have used the primary mass for
the stellar systems if available; we therefore expect the influence of the
companions on our correlations to be small. 
We will also present tests with the subsample of single stars
below. 

\section{Data reduction and analysis}\label{data}

The XEST survey is principally based on CCD camera exposures,
but is complemented with high resolution grating spectra for a few  
bright stars \citep{telleschi06a}, and with Optical
Monitor observations \citep{audard06}. 
The three EPICs onboard {\it XMM-Newton} are CCD-based X-ray cameras that
collect photons from the three telescopes. Two EPIC detectors are of the MOS type
\citep{turner01} and one is of the PN type \citep{strueder01}.
They are sensitive in the energy range of 0.15-15 keV
with a spectral resolving power of E/$\Delta$E = 20-50.

The data were reduced using the Science Analysis System (SAS)
version 6.1. A detailed description of all data reduction procedures
is given in Sect.~4 of \citet{guedel06a}. 

\begin{figure*}
\centering
\includegraphics[angle=-0, width=0.99\textwidth]{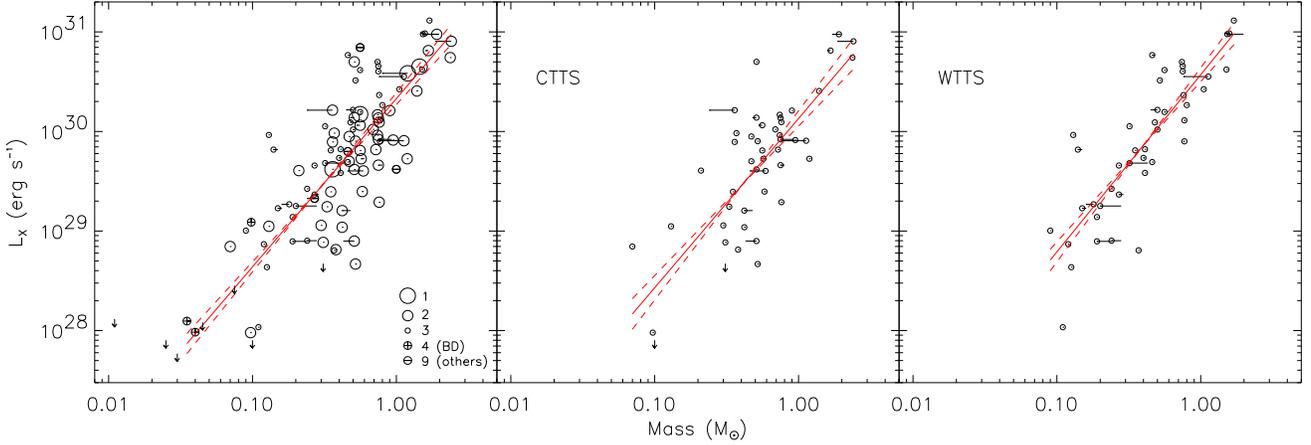}
\caption{X-ray luminosity as a function of mass. From left to right: (a) For all stars.
        (b) For CTTS. (c) For WTTS. The straight lines indicate the regression 
        curves (from the EM algorithm), and the dashed lines illustrate the
        errors in the slopes. }
      \label{mass_lx}
\end{figure*}

Source and background spectra have been obtained for each instrument
using data during the Good Time Intervals (GTIs, i.e., intervals that
do not include flaring background). Further, time intervals with obvious strong 
{\it stellar} flares were 
also excluded from the spectra in order to avoid bias of our results
by episodically heated very hot plasma.

One PN spectrum usually provides more counts than the two MOS spectra together.
We therefore only used the PN data for the spectral analysis,
except for the sources for which PN data were not available (e.g., because
the PN was not operational, or the sources fell into a PN CCD gap).

The spectral fits were performed using two different approaches
in the full energy band. First,
we have fitted the spectra using a conventional one- or 
two-component spectral model (1-$T$ and 2-$T$), both components being 
subject to a common photoelectric absorption. In this approach, the hydrogen
column density, $N_{\rm H}$, two temperatures ($T_{1,2}$) 
and two emission measures (${\rm EM}_{1,2}$) are fitted in XSPEC 
\citep{arnaud96} using the {\it vapec} thermal collisional-ionization 
equilibrium model. 

In the second approach, the spectra were fitted with a model
consisting of a continuous emission measure distribution
(EMD) as found for pre-main sequence and active ZAMS stars
\citep{telleschi05,argiroffi04,garciaalvarez05,scelsi05}.
The model consists of a grid of 20 thermal components
binned to intervals of $d \log T =0.1$ from $\log T= 6$ to $\log T
=7.9$, arranged such that they form an EMD with a peak at a temperature
$T_{\rm 0}$ and two power-laws toward
lower and higher temperatures with power-law indices
$\alpha$ and $\beta$, respectively.
Given the poor sensitivity of CCD spectra at low temperatures,
$\alpha$ was kept fixed at 2, consistent with values
found in previous studies \citep{telleschi05, argiroffi04}, 
while we let $\beta$ free to vary (between $-3 \le \beta \le 1$).
The absorbing hydrogen column density $N_{\rm H}$ was also 
fitted to the data.  The abundances were fixed at values
typical for pre-main sequence stars or very active zero-age 
main-sequence stars \citep{telleschi05,argiroffi04, 
garciaalvarez05, scelsi05}\footnote{The abundance values used
are, with respect to the solar photospheric abundances of \citet{anders89}:
C=0.45, N=0.788, O=0.426, Ne=0.832, Mg=0.263, Al=0.5,
Si=0.309, S=0.417, Ar=0.55, Ca=0.195, Fe=0.195, Ni=0.195}.
For further details, see \citet{guedel06a}.

For each star and each model we computed the average temperature 
($T_{\rm av}$) as the logarithmic average of  all temperatures 
used in the fit, applying the emission measures as  weights.
The X-ray luminosity ($L_{\rm X}$) was computed in the
energy range 0.3-10 keV from the best-fit model
assuming a distance of 140 pc.

Of the 126 members detected in XEST, 22 were 
detected in two different exposures. 
In those cases, two separate spectral fits were made. For correlations
of X-ray parameters with stellar properties, we used logarithmic averages 
of the  results from the two fits. 
On the other hand, if we correlate
X-ray properties with each other, we treat the two spectral
fit results from the same source as different entries.

Results from the spectral fits are given in Table~5 (for the  EMD fits)
and Table~6 (for the 1-$T$ and 2-$T$ fits) of \citet{guedel06a}.
We use the results from the EMD interpretation to perform statistical 
correlations below.

\section{Results}\label{res}

Motivated by results from previous X-ray studies and in particular guided by
the COUP work \citep{preibisch05}, we now seek systematics in the X-ray emission
by correlating X-ray parameters first with fundamental stellar parameters, and
then also seeking correlations among the X-ray parameters themselves. We will consider
the fundamental stellar properties of mass and bolometric luminosity, accretion rate,
and age, but we will not discuss rotation properties here (see \citealt{briggs06}
for a detailed study). The basic X-ray properties used for our correlations
are the X-ray luminosity $L_{\rm X}$ in the 0.3-10~keV band
and the average electron temperature $T_{\rm av}$. One of the main goals of this section is to seek differences
between CTTS and WTTS. We will compare our findings with
those of COUP and some other previous work in Sect.~\ref{discussion}.

\subsection{Correlations between X-rays and stellar parameters}

\subsubsection{Correlation with mass}\label{sect_mass}

In Fig.~\ref{mass_lx} we plot the X-ray luminosity, $L_{\rm X}$ (in erg s$^{-1}$),
 as a function of the stellar mass ($M$, in units of the solar mass, $M_{\odot}$, 
from Table 10 in \citealt{guedel06a}). In the left panel 
we show the relation for all types of objects in our sample.
Different symbols are used to mark different object types (see panel in figure).
Upper limits for non-detections  are marked with arrows.
In the middle and right panels the same relation is shown separately
for CTTS and WTTS, respectively.
A clear correlation is found between the two parameters in all
three plots, in the sense that  $L_{\rm X}$ increases with mass.
The correlation coefficients are 0.79 for the whole sample 
(99 entries), 0.74 for the CTTS (45 entries), and 0.84 for
the WTTS (43 entries).
We  computed the significance of the correlation 
using correlation tests in ASURV (\citealt{lavalley92};  specifically, the 
Cox hazard model, Kendall's tau, and  Spearman's rho have been used) and 
found a probability of $< 0.01$\% that the the parameters are uncorrelated 
in each of the three cases.  As for all subsequent statistical correlation studies,
we summarize these parameters in Table~\ref{summary}.

\begin{table*}
\caption{Summary of results found for the different correlations. In the third column,
         n is the number of stars used in the statistic. $P$ is the probability that the parameters are 
         uncorrelated (computed with ASURV), $C$ is the correlation coefficient and $\sigma$ is the 
         standard deviation from the EM algorithm. The intercept of
         the linear regression is $a$, and $b$ is the slope. Errors are 1-sigma rms values for
         the respective variables.}             
\label{summary}      
\centering                          
\begin{tabular}{l c c rrrr c r r}        
\hline\hline                 
Correlation  & stellar      & n  &  \multicolumn{2}{c}{EM algorithm}  & \multicolumn{2}{c}{bisector algorithm}  &  P & C  & $\sigma$    \\    
             & sample       &    &  \multicolumn{2}{c}{\hrulefill}    & \multicolumn{2}{c}{\hrulefill}          &    &    &          \\     
             &              &    &  $a$\quad\quad\quad & $b$\quad\quad\quad & $a$\quad\quad\quad  &$b$\quad\quad\quad                    &    &    &             \\     
\hline                        
$L_{\rm X}$ vs $M$  & all & 99 & $ 30.33 \pm 0.06$         & $ 1.69 \pm 0.11$&  $ 30.44 \pm 0.05$   & $ 1.91 \pm 0.11$ & $<0.01$\% & 0.79   & 0.45  \\
$L_{\rm X}$ vs $M$  & CTTS & 45 & $ 30.13 \pm 0.09$        & $ 1.70 \pm 0.20$ & $ 30.24 \pm 0.06$   & $ 1.98 \pm 0.20$ & $<0.01$\% & 0.74 & 0.45   \\
$L_{\rm X}$ vs $M$  & WTTS & 43 & $ 30.57 \pm 0.09$        & $ 1.78 \pm 0.17$ & $ 30.69 \pm 0.07$   & $ 2.08 \pm 0.17$ & $<0.01$\% & 0.84 & 0.38   \\
\hline
$T_{\rm av}$ vs $L_{\rm X}$ & CTTS & 19 & $ 6.45 \pm 2.31$ & $ 0.01 \pm 0.08$ & $ -17.95 \pm 13.25$ & $ 0.85 \pm 0.45$ & 43-80\% & 0.06 & 0.22 \\
$T_{\rm av}$ vs $L_{\rm X}$ & WTTS & 29 & $ 2.53 \pm 0.81$ & $ 0.15 \pm 0.03$ & $ 0.13 \pm 0.93$    & $ 0.23 \pm 0.03$ & $<0.01$\% & 0.69& 0.12\\
\hline
$T_{\rm av}$ vs $F_{\rm X}$  & CTTS & 18 & $ 6.77 \pm 0.76$& $ 0.05 \pm 0.12$ & $ 1.67 \pm 1.11$    & $ 0.86 \pm 0.18$ & 62-42\% & 0.11 & 0.22 \\
$T_{\rm av}$ vs $F_{\rm X}$  & WTTS & 32 & $ 5.75 \pm 0.20$& $ 0.18 \pm 0.03$ & $ 5.21 \pm 0.24$    & $ 0.26 \pm 0.03$ & $<0.01$\% & 0.72 & 0.11\\
\hline
$L_{\rm X}$ vs $L_*/L_{\odot}$  & all & 108 & $ 30.00 \pm 0.05$      & $ 1.05 \pm 0.06$ & $ 30.07 \pm 0.04$   & $ 1.11 \pm 0.05$ & $<0.01$\% & 0.83  & 0.44\\
$L_{\rm X}$ vs $L_*/L_{\odot}$  & CTTS & 48 & $ 29.83 \pm 0.06$      & $ 1.16 \pm 0.09$ & $ 29.89 \pm 0.05$  & $ 1.20 \pm 0.10$ & $<0.01$\% & 0.84  & 0.39 \\
$L_{\rm X}$ vs $L_*/L_{\odot}$  & WTTS & 44 & $ 30.22 \pm 0.08$      & $ 1.06 \pm 0.10$ & $ 30.31 \pm 0.06$  & $ 1.25 \pm 0.09$ & $<0.01$\% & 0.85 & 0.41 \\
\hline
$L_{\rm X}/L_{\rm X}(\dot{M})$ vs $\dot{M}$ & CTTS & 37 & $ -4.05 \pm 1.19$ & $ -0.48 \pm 0.15$ & $ -8.32 \pm 1.11$ & $ -1.02 \pm 0.14$ & $0.14-0.52$\% & -0.47 & 0.57 \\
\hline
$L_{\rm X}(M=1 M_{\odot})$ vs age & all & 93 & $ 30.45 \pm 0.06$ & $ -0.36 \pm 0.11$ & $ 30.69 \pm 0.06$ & $ -1.02 \pm 0.07$ & $0.10-0.38$\% & -0.31 & 0.44  \\
\hline
$L_*/L_{\odot}$ vs $M$ & all & 113 & $ 0.23 \pm 0.04$ & $ 1.49 \pm 0.07$ & $ 0.93 \pm 0.04 $ & $1.65 \pm 0.06$ & $<0.01$\% & 0.90 & 0.31  \\
\hline                                   
\hline                                   
\end{tabular}
\end{table*}

We computed linear regression functions for the logarithms of the two parameters, of the
form $\log y = a + b\log x$,
using the parametric estimation  maximization (EM) algorithm in ASURV, which implements the methods
presented by \citet{isobe86}. We find the regression functions $\log L_{\rm X}=(1.69 \pm 0.11)
\log M + (30.33 \pm 0.06)$ for the full stellar sample,
$\log L_{\rm X}=(1.70 \pm 0.20) \log M + (30.13 \pm 0.09)$ for
the CTTS, and $\log L_{\rm X}=(1.78 \pm 0.17) \log M + (30.57 \pm 0.09)$
for the WTTS.  The regression parameters are also listed in Table~\ref{summary}, as for 
all subsequent regression  analyses. 
In the ONC sample, \citet{preibisch05} found the similar linear 
regression $\log L_{\rm X}= (1.44 \pm 0.10) \log M + (30.37 \pm 0.06)$ for all 
stars with masses $< 2 M_{\odot}$ using the same algorithm.

The EM algorithm is an ordinary least-square (OLS) regression of the
dependent variable $y$ ($L_{\rm X}$ in this case) against the 
independent variable $x$ ($M$). When using this method, we assume 
that $L_{\rm X}$ is functionally dependent on the given mass \citep{isobe90}.
However, the $M$ values are also uncertain, and assuming a functional dependence
a priori may not be correct.
We therefore also computed the linear regression using the 
bisector OLS method after \citet{isobe90}, which treats the variables 
symmetrically. In this case, we find $\log L_{\rm X}=(1.91 \pm 0.11)
\log M + (30.44 \pm 0.05)$ for all stars together,
$\log L_{\rm X}=(1.98 \pm 0.20) \log M + (30.24 \pm 0.06)$ for
the CTTS, and $\log L_{\rm X}=(2.08 \pm 0.17) \log M + (30.69 \pm 0.07)$
for the WTTS (see Table~\ref{summary}).
The slopes for the bisector OLS are slightly
steeper than in the EM algorithm. However, the values for
CTTS and WTTS agree within one sigma, and we caution that the 
upper limits for non-detections are not taken into account 
in the bisector linear regression method.

We verified this trend for the subsample of stars
that have not been recognized as multiples.
We find the regression lines $\log L_{\rm X}=(1.72 \pm 0.12)
\log M + (30.39 \pm 0.07)$ using the EM algorithm, and
$\log L_{\rm X}=(1.85 \pm 0.11) \log M + (30.48 \pm 0.07)$ 
using the bisector algorithm. These results are fully
consistent with the results for the total sample.

Differences are present between the CTTS and WTTS stellar samples.
While the slopes found in the correlations for CTTS and WTTS are
consistent within 1$\sigma$, the intercept of WTTS at 1$M_{\odot}
(\log M/M_{\odot}=0)$ is $\approx 0.45$ dex
larger than the intercept for CTTS. This lets us anticipate a larger average
$L_{\rm X}$ in WTTS.
The correlation is better determined for WTTS, as judged from a slightly higher
correlation coefficient and a smaller error in the slope.
Furthermore, the standard deviation, $\sigma$, of the points with respect to 
the regression function from the EM algorithm
is slightly larger for CTTS (0.45) than for WTTS (0.38). 

\subsubsection{Evolution of X-ray emission}

\begin{figure}
\centering
\includegraphics[angle=-0, width=0.49\textwidth]{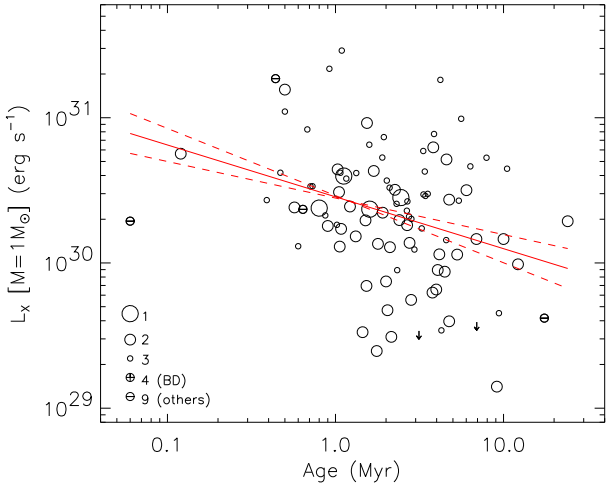}
\caption{Evolution of the normalized X-ray emission. $L_{\rm X}$ has been normalized
         with the predicted values from the $L_{\rm X}$-mass relation 
         (see text for details). The linear regression computed with 
         the EM algorithm is plotted (solid line) together with errors in the slopes 
         (dashed line). Symbols mark different types of stars.}
      \label{age}
\end{figure}

Here, we discuss the evolution of the X-ray emission with
age. Among main-sequence (MS) stars, $L_{\rm X}$ is
correlated with rotation and anti-correlated with
age. The common explanation is that magnetic activity is
directly related to the stellar rotation, and the latter decays with
age because of magnetic braking.
However, TMC PMS stars do not show the relation between $L_{\rm X}$
and rotation observed for MS stars \citep{briggs06}. On the other hand,
$L_{\rm X}$ decreases during the evolution of pre-main sequence stars,
provided that a common X-ray saturation law applies (see below),
because $L_*$ decreases along the Hayashi track.

We have found (see Sect.~\ref{sect_mass}) that $L_{\rm X}$
shows a strong correlation with mass. In order to avoid
an interrelationship between the correlations, we normalize
the measured $L_{\rm X}$ with $L_{\rm X}(M)$ predicted by
the correlation with mass ($L_{\rm X}(M) = 10^{30.33} 
M^{1.69}$~erg s$^{-1}$) and multiply with the  $L_{\rm X}$ expected
for a 1 $M_{\odot}$ star
($10^{30.33}$~erg s$^{-1}$). We designate this quantity
by $L_{\rm X}(M=1M_{\odot})$. 
In Fig.~\ref{age} we plot $L_{\rm X}(M=1M_{\odot})$ as 
a function of age. A slight decline in $L_{X}$ is found
between 0.1 and 10 Myr. The correlation coefficient is 
$C = -0.31$ for 93 entries. The tests in ASURV give
probabilities between 0.1\% and 0.38\% that age 
and $L_{\rm X}(M=1M_{\odot})$ are uncorrelated.
We have computed a linear regression with
the EM algorithm in ASURV and  find $\log(L_{\rm X}/L_{\rm X}(M)) = (-0.36 \pm
0.11) \log~({\rm age}) + (30.45 \pm 0.06)$ (where age is in Myr).
Further, we have tested the linear regression and the correlation
probability when we neglect the two youngest stars
(V410 X4 and LkH$\alpha$ 358) and found the linear regression
to be  consistent within error bars  with the above relation, 
with a probability of $P < 1$\% for no correlation. 
Further, as a test, we have computed the linear regression using the bisector 
algorithm. We find a much steeper slope of $-1.02 \pm 0.07$ 
(Table~\ref{summary}), indicating that the linear regression is nevertheless
only marginal, and the scatter is dominated by other contributions.

\citet{preibisch05b} reported correlations consistent with ours, applying
the EM algorithm to the ONC data.
They correlated $L_{\rm X}$ with age in mass-stratified
subsamples of the surveyed stars and found $L_{\rm X}$ to decrease 
with age with slopes ranging from -0.2 to -0.5, i.e. fully consistent 
with the slope found in Fig.~\ref{age} (mass-stratified analysis for
our sample also indicates decreasing $L_{\rm X}$ in some mass bins but not in others;
our statistics are too poor for this purpose).

\subsubsection{Mass accretion rates}

\begin{figure}
\centering
\includegraphics[angle=-0, width=0.49\textwidth]{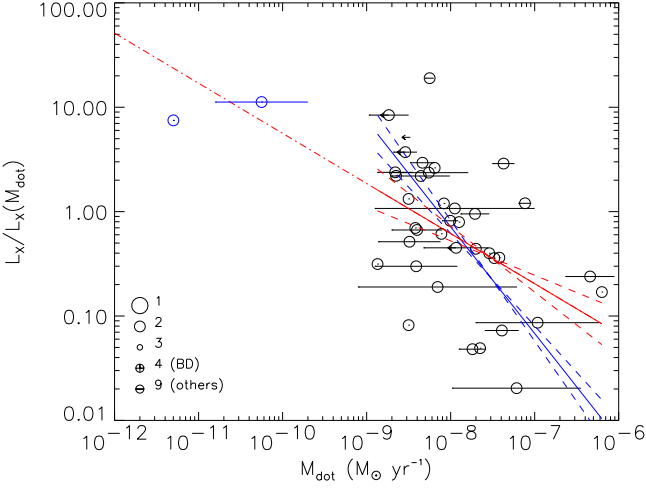}
\caption{Residual X-ray luminosity for CTTS (after normalization with
         the $M-L_{\rm X}$ and $M-\dot{M}$ relation) as a function
         of the mass accretion rate. Regression lines obtained using the 
         EM algorithm (red) and the bisector algorithm (blue) are plotted
         with their respective errors in the slope (dashed lines). The regression lines are computed only 
         using the stars plotted with black circles, while the two stars with
         $\dot{M}$ smaller than $10^{-10} M_{\odot}$~yr$^{-1}$ were ignored (see text for
         more details).}
      \label{mdot_lx}
\end{figure}

\begin{figure}
\centering
\includegraphics[angle=-0, width=0.5\textwidth]{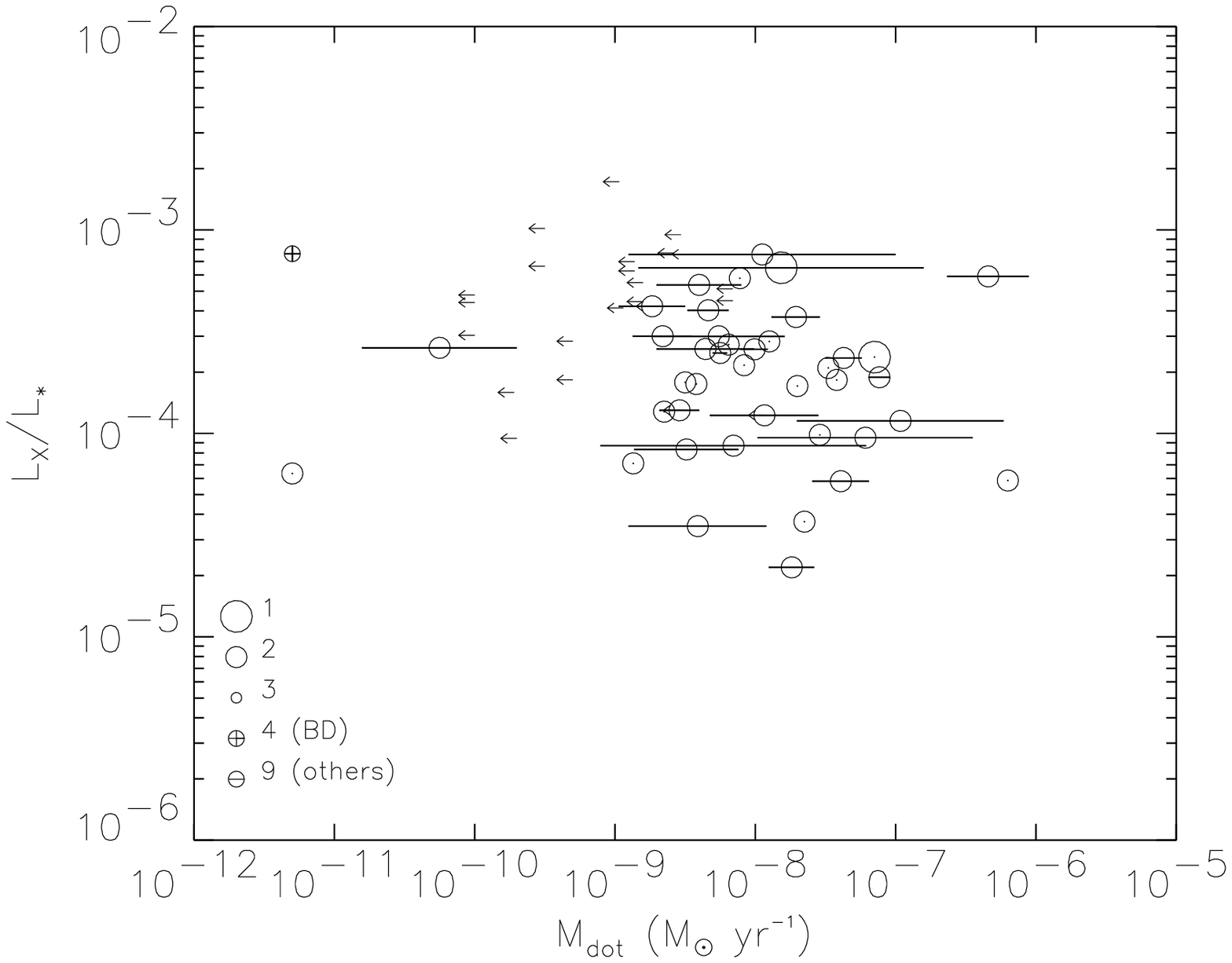}
\includegraphics[angle=-0, width=0.5\textwidth]{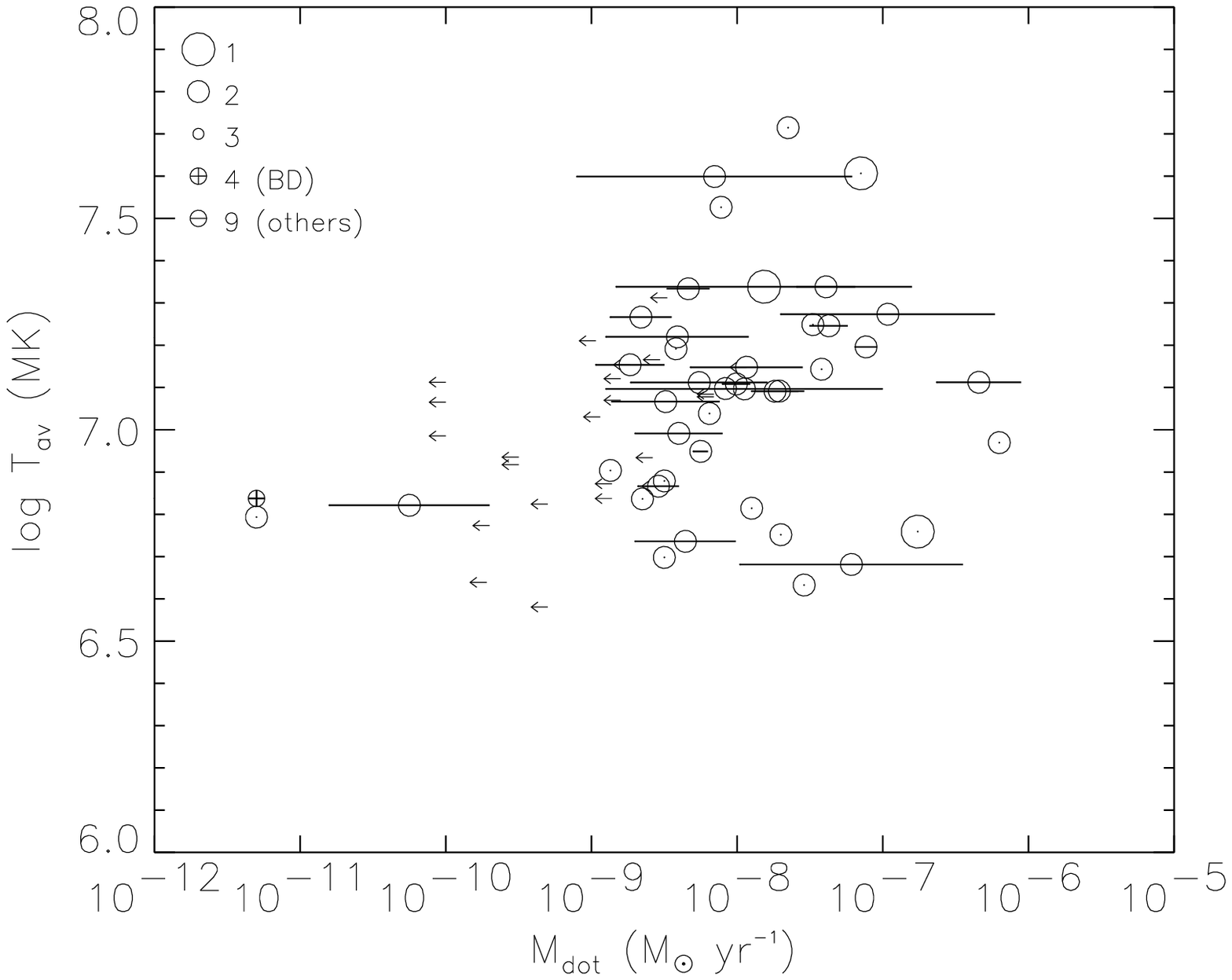}
\caption{Upper panel:
         Fractional X-ray luminosity vs. $\dot{M}$. Lower panel:
         $T_{\rm av}$ vs. $\dot{M}$. Different symbol sizes
         represent different object types as defined 
          in the figures. Arrows represent upper 
         limits for the accretion rates.}
               \label{mdot}
\end{figure}

We now directly compare the X-ray parameters derived from our spectral fits 
($L_{\rm X}$, $L_{\rm X}/L_*$, and $T_{\rm av}$)  with
the previously determined mass accretion rates ($\dot{M}$, in $M_{\odot}$~yr$^{-1}$). We use $\dot{M}$ listed in the
XEST catalog \citep[][and references therein]{guedel06a}. Accretion
rates may be variable, and various methods for their determination may 
produce somewhat different results. If different values were found for
a given star in the literature, the range of $\dot{M}$
is marked by a horizontal line in our figures.

When comparing $L_{\rm X}$ with $\dot{M}$, some caution
is in order. In Sect.~\ref{sect_mass} we have shown that a tight
relation exists between $L_{\rm X}$ and the stellar mass. 
Further, a clear relation between $\dot{M}$ and the stellar mass has
been found for class II objects in the literature 
\citep[e.g.,][]{muzerolle03,muzerolle05,calvet04}. Combining the
$L_{\rm X}$-mass and $\dot{M}$-mass relations, we expect $L_{\rm X}$ to correlate with
$\dot{M}$ as well. However, here we are interested in testing
if an intrinsic relation between the latter two parameters exists that
is not a consequence of the two former relations.
\citet{calvet04} have used evolutionary tracks of \citet{siess00} 
(consistent with the XEST survey) to find a relation $\dot{M} 
\propto M^{1.95}$ in the mass range between 0.02 and 3 $M_{\odot}$.
Similarly, \citet{muzerolle03} and \citet{muzerolle05} found 
$\dot{M} \propto M^{2}$ and  $\dot{M} \propto M^{2.1}$ respectively, using
different evolutionary tracks. We therefore adopt the relation
$\log \dot{M} \approx 2 \log M - 7.5$. Further, we use $\log L_{\rm X} =
1.69 \log M + 30.33 $ (Sect.~\ref{sect_mass}), and  we then compute the 
expected $L_{\rm X}$  for each $\dot{M}$ value, namely $\log L_{\rm X} (\dot{M})
= 0.85 \log \dot{M} + 36.67$.

In Fig.~\ref{mdot_lx} we plot the ratio of $L_{\rm X}/L_{\rm X}(\dot{M})$
as a function of $\dot{M}$ for class II objects.
We will expect that the values scatter around a constant if this ratio were determined 
only by the $M-L_{\rm X}$ and $M - \dot{M}$ relations.
We find a very large scatter for any given $\dot{M}$ (2--3 orders of magnitude) 
but, using a regression analysis, a tendency for weak accretors to 
show higher $L_{\rm X}$, compared to strong accretors.
However, if we exclude the two stars with the smallest accretion 
rates (plotted with blue symbols in Fig.~\ref{mdot_lx}), the correlation 
is less clear. In this case, the correlation coefficient  
is $C=-0.47$ for 37 data points. Nevertheless, the probability, computed in ASURV for 
the EM algorithm, that {\it no} correlation is present is only $ P = $~0.1\% -- 0.5\%. 
We have computed the linear regression using different methods.
With the EM algorithm we find
$\log L_{\rm X}/L_{\rm X}(\dot{M}) = (-0.48 \pm 0.15) \log \dot{M} - 
(4.05 \pm 1.19)$ (Table~\ref{summary}). Using the bisector algorithm, 
however, the slope is found to be $-1.02 \pm 0.14$, i.e., more
than 3 sigma steeper than the slope found with the EM algorithm.
The entries for two stars with low $\dot{M}$ (plotted in blue) are
consistent with the linear regression found with the EM algorithm. We conclude that 
the two parameters are not evenly distributed, but that a linear regression of the
logarithmic values cannot clearly be claimed.

In Fig.~\ref{mdot} we plot $L_{\rm X}/L_*$ and  $T_{\rm av}$
as a function of the accretion rate. 
Arrows represent upper limits for $\dot{M}$.
In both cases no  correlation is evident.

\subsubsection{Correlation with bolometric luminosity}\label{sect_lbol} 

\begin{figure*}
\centering
\includegraphics[angle=-0, width=1.02\textwidth]{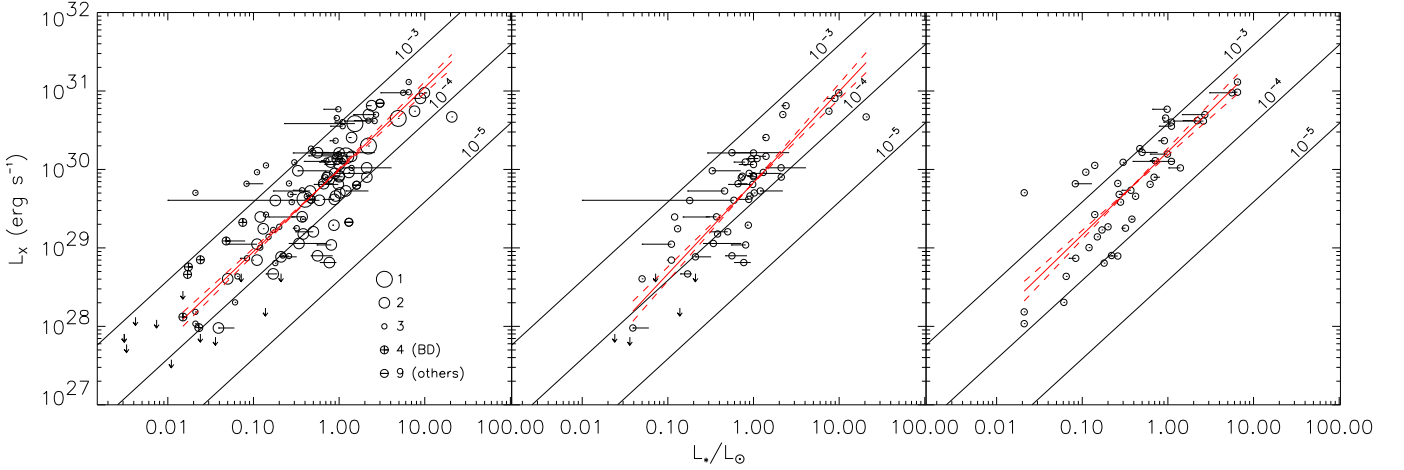}
\caption{$L_{\rm X}$ as a function of the $L_*$. Left: for all stars;
         the different symbols describe different classes of stars, 
         while arrows are upper limits for non-detections. Middle: 
         same for CTTS (type 2). Right: same for WTTS (type  3). The horizontal bars
         show the ranges of literature values for $L_*$. }
      \label{lx_lbol}
\end{figure*}

In Fig.~\ref{lx_lbol} we plot $L_{\rm X}$ 
as a function of the stellar bolometric luminosity $L_*$ (from 
\citealt{guedel06a} and references therein). The lines corresponding to
$L_{\rm X}/L_* = 10^{-3}, 10^{-4}$, and $10^{-5}$ are also shown.
In the left panel, all stars are plotted, with
different symbols for each stellar class as described
in the figure. We again excluded from the plot the stars mentioned
in Sect.~\ref{sample}. 
Upper limits for non-detections are marked with arrows. 
By far most of the stars are located between
$L_{\rm X}/L_* = 10^{-3}$ and $L_{\rm X}/L_* = 10^{-4}$.
In the middle and right panels, we present CTTS and WTTS
separately.

The correlation coefficients are 0.83 for the full stellar sample (108 entries), 
and 0.84 (48 entries) and 0.85 (44 entries) for CTTS and WTTS, respectively.
Probabilities for the absence of a correlation are very small, $P < 0.01$\%.
We computed linear regression lines with the EM algorithm in 
ASURV. For the full sample, we find $\log L_{\rm X} = (1.05 \pm 
0.06) \log L_*/L_{\odot} + (30.00 \pm 0.05)$, whereas for CTTS and WTTS,
$\log L_{\rm X} = (1.16 \pm 0.09) \log L_*/L_{\odot} + (29.83 \pm 0.06)$
and $\log L_{\rm X} = (1.06 \pm 0.10) \log L_*/L_{\odot} + (30.22 \pm 0.08)$,
respectively. Fig.~\ref{lx_lbol}c shows one WTTS at $L_*/L_{\odot} \approx 0.01$
with a rather high $L_{\rm X} \approx 5\times 10^{29}$~erg~s$^{-1}$ (KPNO-Tau 8 = XEST-09-022).
Not considering this object, the slope of the regression slightly steepens to
$1.17\pm 0.09$, which is only marginally different from the slope based on all WTTS. The 
standard deviation at the same time marginally decreases from 0.41 to 0.36.
 
Again, we also computed a linear regression
using the bisector OLS algorithm that treats both $L_*$ and $L_{\rm X}$
as independent variables. The slopes are very similar to the those 
reported above (Table \ref{summary}).
The important distinction between CTTS and WTTS is that the latter clearly 
tend to be located at higher $L_{\rm X}/L_*$ (see also below):
at $L_*/L_{\odot} = 1 $ the CTTS show an average $\log L_{\rm X} = 29.83$
~[erg~s$^{-1}$], while for the WTTS, $\log L_{\rm X} = 30.22$~[erg~s$^{-1}$].

The regressions are thus compatible with a linear relation between $L_{\rm X}$
and $L_*$, and therefore $L_{\rm X}/L_*$ is, on average for a given $L_*$, a constant
between $10^{-4}$ and $10^{-3}$ regardless of $L_*$. This is reminiscent of the situation among
very active, rapidly rotating main-sequence or evolved subgiant stars that {\it
saturate} at fractional X-ray luminosities of the same order, provided they rotate sufficiently
rapidly. We thus find that the majority of our TTS are in a saturated state. A consequence
of this would be that rotation no longer controls the X-ray output, as suggested 
by \citet{preibisch05} for the Orion sample. This is discussed for the XEST sample
by  \citet{briggs06}. Below, we will specifically
study whether the $L_{\rm X}/L_*$ relation is different for CTTS and for WTTS.

The correlation found for the full stellar sample
is compatible with the relation found in Orion:
$\log L_{\rm X} = (1.04 \pm 0.06) \log (L_*/L_{\odot}) + 
(30.00 \pm 0.04)$  \citep{preibisch05}.
The slope found for the WTTS in the ONC 
is also consistent with our results within the error bars. 
For CTTS, on the other hand, \citet{preibisch05} found a very large scatter in the correlation. 
This is not observed in our XEST sample; we rather see similar scatter for CTTS and WTTS, as 
demonstrated by the similar correlation coefficients, the similar errors in the slope,
and the similar standard deviations. \citet{preibisch05}
suggested that strong accretion could lead to larger errors 
in the determination of stellar luminosity  and the effective temperature.

\begin{figure}
\centering
\includegraphics[angle=-0, width=0.48\textwidth]{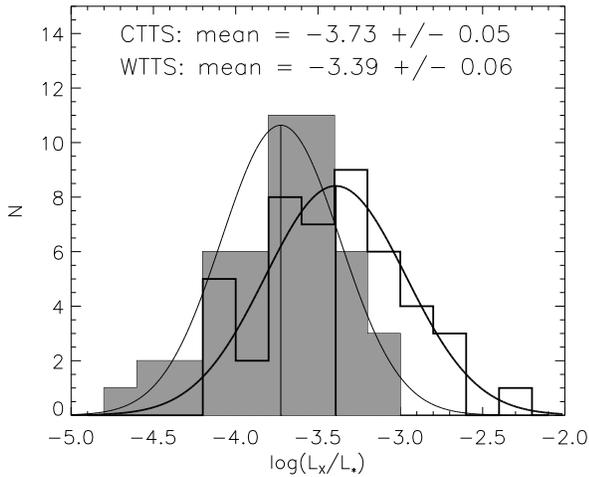}
\caption{Distributions of $\log (L_{\rm X}/L_*)$ for CTTS (grey 
         histogram) and WTTS (white histogram). Gaussians that fit the distributions 
         are also plotted and their peaks are marked with vertical lines.}
      \label{lxlbol_hist}
\end{figure}

\begin{figure}
\centering
\includegraphics[angle=-0, width=0.48\textwidth]{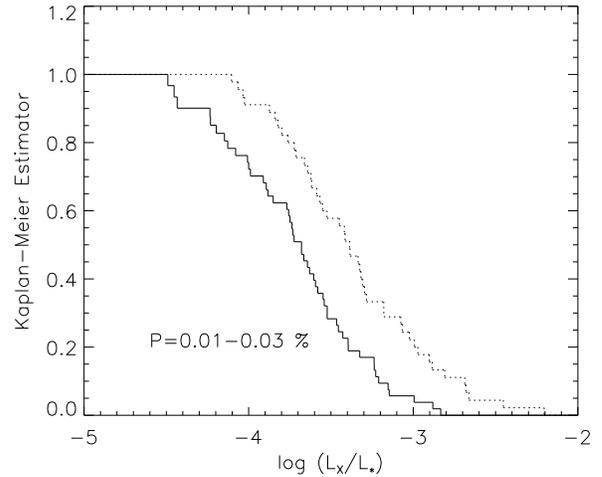}
\caption{Cumulative distribution of $\log (L_{\rm X}/L_*)$ for CTTS (solid) and for WTTS (dotted). }
      \label{lxlbol}
\end{figure}

\begin{figure*}
\centering
\includegraphics[angle=-0, width=0.98\textwidth]{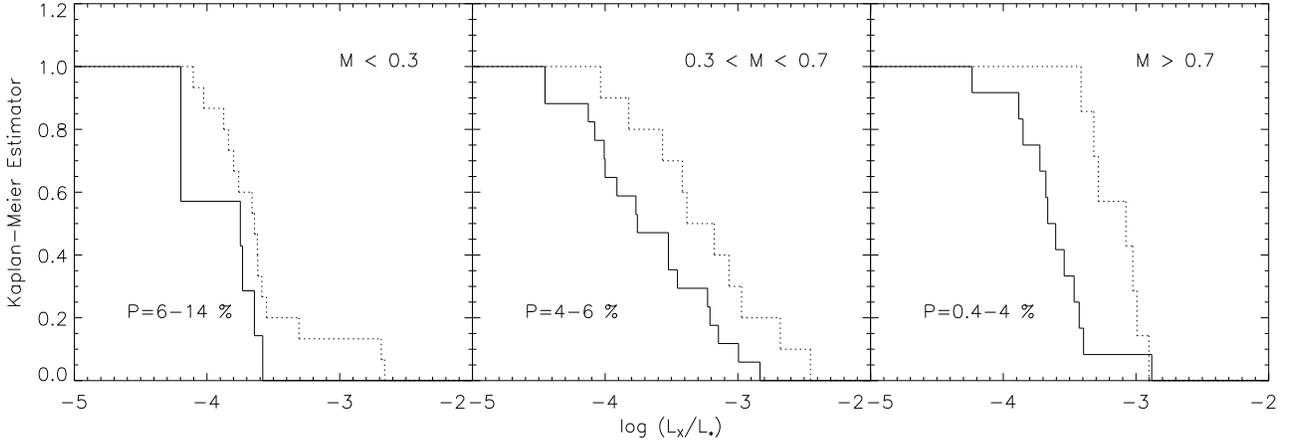}
\caption{Cumulative distributions of $\log (L_{\rm X}/L_*)$ for WTTS  (dotted) and CTTS (solid)
         for masses smaller than 0.3 $M_{\odot}$ (left), masses between 
         0.3 and 0.7 $M_{\odot}$ (middle), and masses larger than 0.7 
         $M_{\odot}$ (right). }
      \label{lxlbol3}
\end{figure*}


\subsubsection{Fractional X-ray luminosity $L_{\rm X}/L_*$}{\label{lxlbolsect}}

In Fig.~\ref{lxlbol_hist} we plot the histogram for the distribution of $\log
(L_{\rm X}/L_*)$ for CTTS (grey) and WTTS (white). 
The two populations are different, with WTTS having a larger 
mean $\log (L_{\rm X}/L_*)$. We fitted each of the two histograms with 
a Gaussian function and computed the mean and its errors. For CTTS, we 
find  $\langle \log (L_{\rm X}/L_*) \rangle = -3.73 \pm 0.05$, while 
for WTTS, $\langle \log (L_{\rm X}/L_*) \rangle = -3.39 \pm 0.06$.

A more rigorous test is based on the Kaplan-Meier 
estimators as computed in ASURV, which implements
the methods presented by \citet{feigelson85}. This method also accounts for
the upper limits in $L_{\rm X}$ for the non-detections. The results are plotted in 
Fig.~\ref{lxlbol}. The solid line represents the CTTS, the dotted 
line the WTTS.
The WTTS distribution is shifted toward larger $\log (L_{\rm X}/L_*)$
compared to the CTTS distribution by a factor of approximately  2.
We find $\langle \log (L_{\rm X}/L_*) \rangle = -3.72 \pm 0.06$
and $\langle \log (L_{\rm X}/L_*) \rangle = -3.36 \pm 0.07$ for CTTS and WTTS, 
respectively, in full agreement with the Gaussian fit.
Judged from a two-sample test based on the Wilcoxon test
and logrank test in ASURV,
the probability that the two distributions are obtained
from the same parent population is very low, namely $P = 0.01$\%$-0.03$\%.

Again, we test this result using the subsample of stars that have not
been recognized as multiples. The subsample consists on 29 CTTS 
(4 of which  have upper limits) and 33 WTTS (with no upper limits). We find a probability of  
$P = 0.05$\%$-0.07$\% that the distributions arise from
the same parent population, and $\langle \log (L_{\rm X}/L_*) 
\rangle_{\rm CTTS} = -3.83 \pm 0.06$ and $\langle \log (L_{\rm X}/L_*) 
\rangle_{\rm WTTS} = -3.40 \pm 0.08$. These results are consistent 
with the results found in the full sample. We can therefore conclude 
that multiplicity does not influence our results.

In Fig.~\ref{lxlbol3} we plot the Kaplan-Meier estimator for the
distribution of $\log (L_{\rm X}/L_*)$ in three different mass ranges:
for stars with masses smaller than 0.3 $M_{\odot}$, between 0.3 and 0.7 
$M_{\odot}$, and larger than 0.7 $M_{\odot}$.
For the latter two mass bins,
the distributions belong to two different parent populations
at the $> 94$\% level. For lower masses ($M < 0.3 M_{\odot}$),
we find the CTTS and WTTS distributions to be similar, but we still
find  larger $\log (L_{\rm X}/L_*)$ for WTTS than than for CTTS. 
The difference between the two populations is not significant at the $\approx$ 10\% level
possibly because of the small size of the stellar sample in 
this mass range.

Our results can be compared with the distributions found in the COUP 
survey, shown in  Fig.~16 of  \citet{preibisch05}. In the latter figure, 
the stars are classified according to the 8542 \AA\ Ca II line, which 
is an indicator of disk accretion, similar to the EW(H$\alpha$) 
used in our work. In Orion, a substantial difference has  been found 
between the distributions of accreting and non-accreting stars
in the mass ranges 0.2--0.3 $M_{\odot}$ and 0.3--0.5 $M_{\odot}$.
However, for 0.5--1 $M_{\odot}$, the two distributions appeared to be
compatible, in contrast to our findings that show fainter CTTS consistently in
all mass ranges.

\begin{figure*}
\centering
\includegraphics[angle=-0, width=0.98\textwidth]{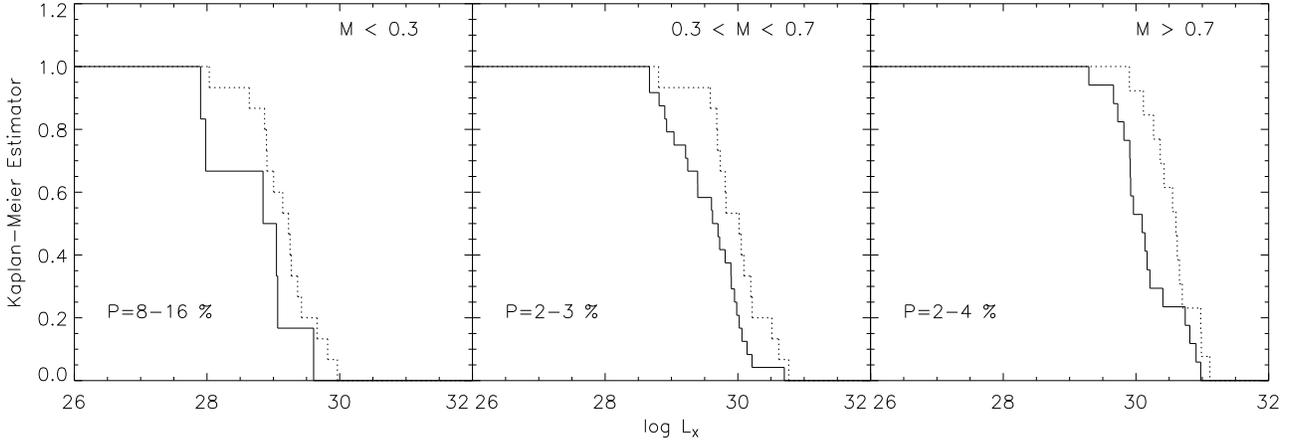}
\caption{X-ray Luminosity Function (XLF) for CTTS (solid) and WTTS (dotted) for 
     for masses smaller than 0.3 $M_{\odot}$ (left), masses between 
         0.3 and 0.7 $M_{\odot}$ (middle), and masses larger than 0.7 
         $M_{\odot}$ (right).}
      \label{xlf3}
\end{figure*}

\begin{figure}
\centering
\includegraphics[angle=-0, width=0.48\textwidth]{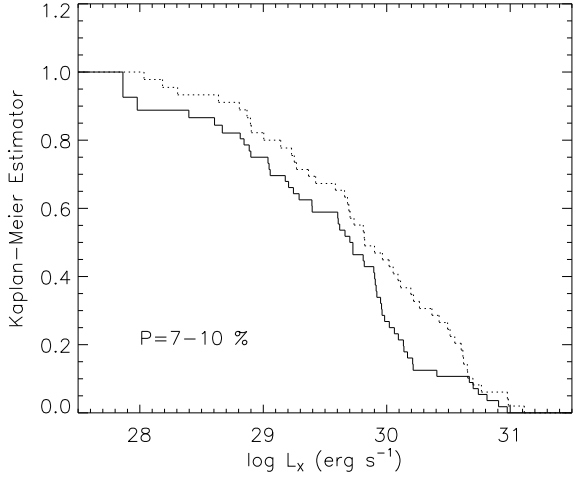}
\caption{X-ray Luminosity Function (XLF) for CTTS (solid) and  WTTS (dotted).}
      \label{xlf}
\end{figure}

\subsection{Correlations between X-ray parameters}

\subsubsection{The X-ray luminosity function}\label{sec_xlf}

In Fig.~\ref{xlf} we display the X-ray luminosity function (XLF)
for WTTS and CTTS for our Taurus sample. The XLF has again
been calculated using the Kaplan-Meier estimator in ASURV,
so that the few upper limits have also been considered.  
The total number of sources used was 105, 56 of them being
CTTS (including 6 upper limits) and 49 WTTS (including 1 upper limit). 
The WTTS are again more luminous than CTTS by a factor of about 2
(with mean values $\langle \log L_{\rm X} \rangle_{\rm C}=29.51$
and $\langle \log L_{\rm X} \rangle_{\rm W}=29.80$). 
The probability that the two distributions arise from the same 
parent population is $7$\%$-10$\%, computed using ASURV as described
in Sect.~\ref{lxlbolsect}.

If we restrict the stellar sample to stars with no recognized 
multiplicity, we obtain average X-ray luminosities of $\langle 
\log L_{\rm X} \rangle_{\rm C}=29.38$ and $\langle \log L_{\rm 
X} \rangle_{\rm W}=29.65$, for samples consisting of 32 CTTS (5 
upper limits) and 36 WTTS (1 upper limit). 
The difference between the two stellar samples is 0.3 dex (i.e., a factor
of two), similar to what we found for the full sample. However, the two-sample
tests give a larger probability (P = 12\%--33\%) that the two stellar groups
arise from the same parent population. Among the multiple sources, 24
are CTTS (with 1 upper limit), but only 13 are WTTS (with no upper limit).
By adding the multiples to the sample of single stars, we expect that the distributions
slightly shift toward larger $L_{\rm X}$, and because there are significantly
more multiple CTTS, the CTTS distribution of the total sample should be
more similar to the WTTS total distribution, but the opposite trend is seen.
We conclude that the trends are grossly the same for the total sample and the
single-star subsamples, the larger probability being due to significantly smaller
samples that are compared. Overall, thus, CTTS are recognized as being X-ray deficient
when compared to WTTS. 

Fig.~\ref{xlf3} shows the X-ray luminosity function for
the same three mass ranges as used in Sect.~\ref{lxlbolsect} ($M
< 0.3 M_{\odot}$ in the left panel, $0.3 M_{\odot} < M < 0.7 
M_{\odot}$ in the middle panel, and $M > 0.7_{\odot}$ in the right panel).
Again, we find the largest difference between CTTS and WTTS
for the two higher-mass bins, with probabilities
of only $2$\%$-4$\% that the distributions belong to the same 
parent population. The probability is substantially larger 
for $M < 0.3 M_{\odot}$ ($29$\%$-32$\%), but the statistics are also considerably poorer.

\begin{figure}
\centering
\includegraphics[angle=-0, width=0.48\textwidth]{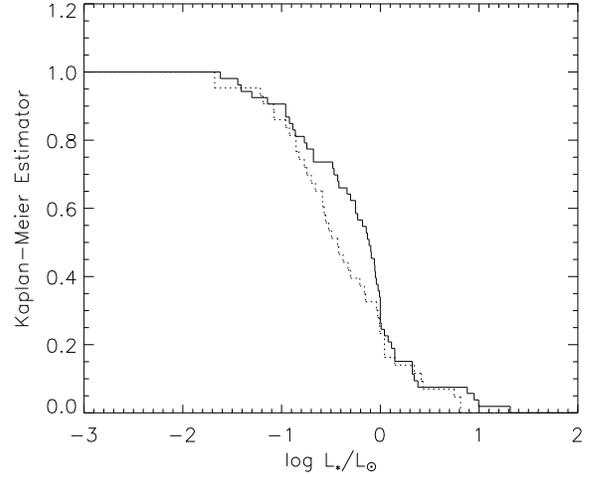}
\caption{Stellar bolometric luminosity function for CTTS (solid) and WTTS (dotted). }
      \label{lbol}
\end{figure}

Considering the difference in the XLFs of CTTS and WTTS alone, a possible cause
could be that the bolometric luminosity function of CTTS would indicate lower 
luminosities $L_*$ than for WTTS, which would result in lower average $L_{\rm X}$ 
provided that  $L_{\rm X}/L_* \approx $ constant, i.e., that saturation applies for 
all stars. In Fig.~\ref{lbol}, we plot the distributions of $L_*$
for WTTS and CTTS. In fact, the CTTS are found to be slightly {\it more luminous} than the
WTTS. We find  $\langle \log L_* \rangle_C=33.35 \pm 0.08$  
and  $\langle \log L_* \rangle_W=33.19 \pm 0.09$. The probability that the
distributions arise from the same parent population is $15$\%$-21$\%, i.e.
making the difference marginal. We conclude that because $L_{\rm X}$ is linearly 
correlated with $L_*$ (Fig.~\ref{lx_lbol}), 
the difference in $L_{\rm X}$ for the two samples is intrinsic, which is of course 
a reconfirmation of our previous finding that the distributions of  
$L_{\rm X}/L_*$  also indicate lower activity for CTTS compared to WTTS.

\subsubsection{Absorption}

\begin{figure}
\centering
\includegraphics[angle=-0, width=0.49\textwidth]{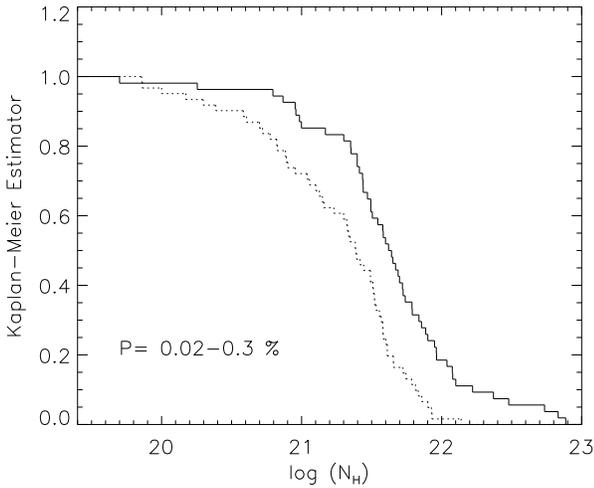}
\caption{Cumulative distribution of $N_{\rm H}$ for
CTTS (solid line) and WTTS (dotted line).}
\label{fig:nh}
\end{figure}

In Fig.~\ref{fig:nh} we plot the distribution of $N_{\rm H}$ 
for accreting (solid line) and non-accreting stars (dotted
line) calculated using the Kaplan-Meier estimator in ASURV. 
The logarithmic average of $N_{\rm H}$ for CTTS is
$\langle N_{\rm H} \rangle_C= 4.2 \times 10^{21}$~cm$^{-2}$  
and is more than a factor of two larger than the average for WTTS 
($\langle N_{\rm H} \rangle_W= 1.8 \times 10^{21}$~cm$^{-2}$).

The $N_{\rm H}$ values found from our spectral fits are
roughly consistent with the visual extinctions $A_{\rm V}$ and the
infrared extinctions $A_{\rm J}$ if we assume
a standard gas-to dust ratio ($ N_{\rm H}/A_{\rm V} = 2 \times 
10^{21}~{\rm cm}^{-2}~{\rm mag}^{-1}$, $N_{\rm H}/A_{\rm J} = 7.1 
\times 10^{21}~{\rm cm}^{-2}~{\rm mag}^{-1}$; \citealt{vuong03}).
For a detailed discussion on the gas-to-dust ratio in
TMC, we refer the reader to Glauser et al. (2007, in preparation).
High photoelectric absorption might influence the spectral fits
to low-resolution spectra, because the coolest plasma components,
more affected by absorption, cannot be reliably quantified.

\begin{figure}
\centering
\includegraphics[angle=-0, width=0.49\textwidth]{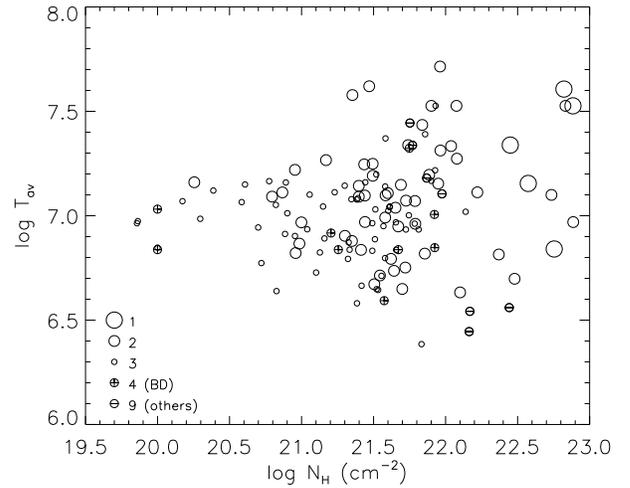}
\caption{$T_{\rm av}$ as a function of $N_{\rm H}$ for our sample.}
\label{nh_tav}
\end{figure}

We studied possible biases introduced by high absorption
by correlating  $N_{\rm H}$ with $T_{\rm av}$ and $L_{\rm X}$.
We have found $L_{\rm X}$ to range between approximately
$10^{28}$ erg s$^{-1}$ and $10^{31}$ erg s$^{-1}$, independent
of the photoelectric absorption.
The uncertainty of the determination of $L_{\rm X}$, on the other hand, does 
increase with increasing $N_{\rm H}$ (and decreasing number of
counts in the spectrum) as derived by \citet{guedel06a}.
Similarly, if we correlate  $T_{\rm av}$ with $N_{\rm H}$ 
(Fig.~\ref{nh_tav}),  we find a larger range of $T_{\rm av}$ 
(symmetrically around $\log T_{\rm av} \approx 7.0$ [K]) for
highly absorbed sources, while $T_{\rm av}$ is found
to be similar for all the sources with low absorption ($N_{\rm H}
< 10^{21}$ cm$^{-2}$), $\log T_{\rm av} \approx 7.0 \pm 0.2$~[K].
The larger range of $T_{\rm av}$ at higher $N_{\rm H}$ is likely to be
the result of larger scatter due to less reliable spectral fitting.
In any case, there is no trend toward higher $T_{\rm av}$ for higher 
$N_{\rm H}$.

\subsubsection{Correlation of $L_{\rm X}$ with electron temperature}\label{sect_temp}

\begin{figure*}
\centering
\includegraphics[angle=-0, width=0.80\textwidth]{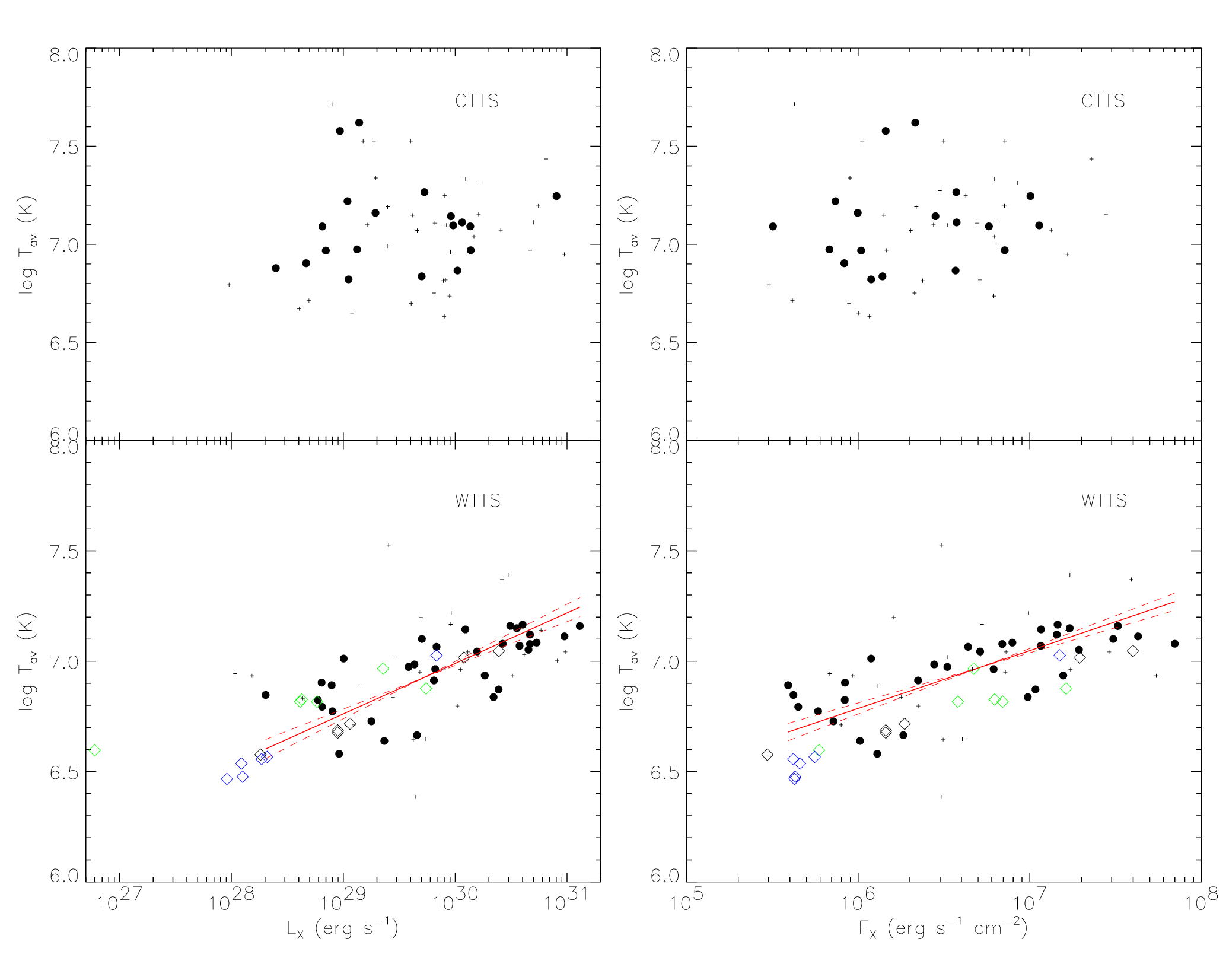}
\vskip -0.6truecm
\caption{Left pannel: $T_{\rm av}$ as a function of $L_{\rm X}$ for CTTS
(top) and WTTS (bottom). Right panel: $T_{\rm av}$ as a function of $F_{\rm X}$ for CTTS
(top) and WTTS (bottom). The low-absorption TTS samples are marked by filled black
bullets, while small crosses give loci of high-absorption objects or sources with few
counts (see text for details). Black diamonds mark solar analog stars \citep{telleschi05} and  
blue and green diamonds mark K- and M-type main-sequence stars, respectively (see text for
references). The straight lines in the WTTS plots are linear regression fits 
(based on bisector regression, the dashed lines illustrating the error ranges in the slopes).}
      \label{lx_tav}
\end{figure*}

In Fig.~\ref{lx_tav} we plot $T_{\rm av}$ as a function of $L_{\rm X}$
and as a function of the X-ray surface flux ($F_{\rm X}$)
for CTTS and WTTS, respectively. The surface fluxes have been calculated 
using the radii reported in Table 10 of \citet{guedel06a}. 
In the plots for the WTTS we also show values for six main-sequence G-type solar 
analog  stars \citep{telleschi05}, for 5  K-type stars (AB Dor from 
\citealt{sanz03},  and $\epsilon$ Eri,
70 Oph A\&B, 36 Oph A\&B from \citealt{wood06}), and  for 6 M-type 
main-sequence stars (EQ Peg, AT Mic,
AD Leo and EV Lac from \citealt{robrade05}, AU Mic from \citealt{magee03}, and
Proxima Cen from \citealt{guedel04a}).

Spectral fits to low-resolution spectra that are subject to photoelectric
absorption tend to ignore the coolest plasma components, as the soft part
of the spectrum is most severely affected by the absorption. 
CTTS are on average more absorbed than WTTS. Given the larger absorptions, 
there could be a bias toward higher average temperatures in CTTS,
although such a trend is not visible in Fig.~\ref{nh_tav}.
We nevertheless counteract a possible residual
bias by restricting the stellar sample used for the correlation to stars with
$N_{\rm H} < 3\times 10^{21}$~cm$^{-2}$. The logarithmic means of $N_{\rm H}$
for CTTS and WTTS after these restrictions are $1.2 \times 10^{21}$~cm$^{-2}$ and
$8\times 10^{20}$~cm$^{-2}$, respectively, making these samples very similar 
with regard to absorption properties.  

Further, we exclude the very faint sources
(with less than 100 counts collectively in the three detectors) that could
also produce unreliable $L_{\rm X}$ and $T_{\rm av}$ results.
In Fig.~\ref{lx_tav} the filled circles represent the stars used for the linear regression fit, 
while the stars excluded from the fit are plotted with small crosses. 
Overall, the absorbed and faint sources fit well to the
trends found from less absorbed and more luminous sources,
but their scatter tends to be larger.

For CTTS, we find almost no correlation between $T_{\rm av}$ and 
$L_{\rm X}$ or $F_{\rm X}$ (the correlation coefficients are 0.06 
and 0.11, respectively). On the contrary, for WTTS  $T_{\rm av}$ is clearly
correlated with both   $L_{\rm X}$ and $F_{\rm X}$. The correlation coefficients 
are 0.69 and 0.72 for $L_{\rm X}$ and $F_{\rm X}$, with 33 and 32 data points, 
respectively (Table~\ref{summary}).
The probability that no correlation is present is $< 1$\%
in either case. We computed the linear regression using the bisector
OLS algorithm (no a priori relation between the two measured variables assumed)
to find
$\log T_{\rm av} = (0.23 \pm 0.03) \log L_{\rm X} + (0.13 \pm 0.93)$
and 
$\log T_{\rm av} = (0.26 \pm 0.03) \log F_{\rm X} + (5.21 \pm 0.24)$.
WTTS follow a trend similar to that shown by MS stars in the $T_{\rm av}$
vs. $L_{\rm X}$ relation. In the $T_{\rm av}$ vs. $F_{\rm X}$ relation, on the 
other hand, we find that WTTS are in general hotter than MS stars for
a given $F_{\rm X}$.

We have checked these results using the EM algorithm, finding slightly 
shallower slopes (see Table~\ref{summary}). Shallower
slopes are expected in the EM algorithm when compared to the bisector 
OLS algorithm \citep{isobe90}. For CTTS, 
where no correlation is found, the two algorithms result in completely different
slopes (Table~\ref{summary}), an indication of absence of a linear regression 
(\citealt{isobe90}; the different slopes in the absence of a correlation are
a consequence of the defining minimization of the algorithm. The EM algorithm 
returns a slope of $\approx 0$, whereas the bisector algorithm yields a slope
around unity).

\begin{figure}
\centering
\includegraphics[angle=-0, width=0.49\textwidth]{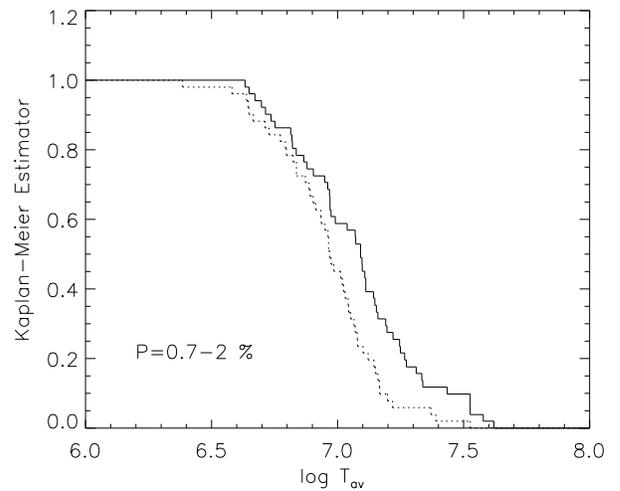}
\caption{Kaplan-Meier estimator for $T_{\rm av}$ for CTTS (solid)
          and WTTS (dotted). Only sources with more than 100
         counts in the three EPIC detectors are used.}
      \label{tav_kaplan}
\end{figure}

We use the Kaplan-Meier estimator in ASURV to compare the distributions
of $T_{\rm av}$ for CTTS and WTTS. The two distributions are shown in 
Fig.~\ref{tav_kaplan} for the case where we do not apply a restriction to $N_{\rm H}$: 
the solid line represents CTTS, while the dotted line
represents WTTS. Only stars with more than 100 counts in the three EPIC
detectors are used. 
The probability that the two distributions arise from the same parent population is 
$0.7$\%$-2$\%. If we restrict the sample to stars
with low $N_{\rm H}$ ($< 3 \times 10^{21}$~cm$^{-2}$), we find
the mean value of $\log T_{\rm av}$[MK] $ = 7.10$ [MK] with $\sigma=0.22$ 
for CTTS and $\log T_{\rm av}$[MK] $ = 6.88$ with $\sigma=0.17$ for WTTS.
The distribution is similar to the one found in Fig.~\ref{tav_kaplan}
and the probability for the CTTS and WTTS distributions
to originate from the same parent population is again only $0.7$\%$-2$\%.

We have checked if the difference found
in the plasma temperatures of WTTS and CTTS could be attributed to 
abundance anomalies that may not have correctly been accounted for in the 
spectral fits. \citet{kastner02}, \citet{stelzer04}, and \citet{drake05} 
found large Ne/Fe and Ne/O abundance ratios in the spectrum of TW Hya.
We have therefore fitted the spectra of  the 19 CTTS 
with $N_H < 3 \times 10^{21}$~cm$^{-2}$ and more than 100 counts in the
combined EPIC spectra (filled black bullets in Fig.~\ref{lx_tav})
adopting abundances as found in TW Hya (O = 0.2, Ne = 2.0, Fe = 0.2, with
respect to the solar photospheric abundances of \citealt{anders89}, all 
other abundances as given in Sect.~\ref{data}). 
The average temperatures obtained with this model are generally consistent within 0.1 dex 
with the temperatures found based on our standard abundances. 
Only for one star, HO Tau AB, did we find an average temperature
significantly lower, while the 
general trend toward higher $T_{\rm av}$ for CTTS remains unchanged.  
We can therefore
exclude that the difference in temperatures is induced by abundance
anomalies as those observed in TW Hya.

\citet{telleschi06a} derived the thermal structure of nine
pre-main sequence stars from XEST based on high-resolution
Reflection Grating Spectrometer data, using variable abundances.
They found $T_{\rm av}$ to be compatible with values used here, which 
were derived from EPIC CCD spectra (an exception is the CTTS SU Aur,
for which the temperature found with RGS is even higher than that derived from
the EPIC spectra). 
In the latter work, however, a difference in the abundances has been found between stars
of spectral type K and stars of spectral type G. The abundances
found for the K-type stars reflect approximately the abundances used for 
the XEST EPIC fits (following an inverse FIP effect), while G-type stars
show lower Ne/Fe and O/Fe abundance ratios. We therefore fitted the four
G-type stars in our stellar sample with an abundance pattern as
found for this spectral class by \citet{telleschi06a}. Again,
we did not find a significant change in temperatures.
In summary, we do not
find any appreciable effect that abundance anomalies other than
those adopted in our study might have on the temperature determination.
We also note that Scelsi et al. (2006, in preparation) 
studied the abundances derived from the EPIC spectra of the brightest 
sources in the XEST sample and found average abundances very similar to 
the standard abundances used in our CCD fits.

The difference in the coronal temperatures of CTTS and WTTS is
in particular due to the larger $T_{\rm av}$ found in the CTTS with low 
$L_{\rm X}$. We therefore calculated the mean of all $T_{\rm av}$ values for
CTTS and WTTS with $L_{\rm X} < 3 \times 10^{29}$~erg s$^{-1}$,
low absorption ($N_{\rm H} < 3\times 10^{21}$~cm$^{-2}$), and more than
100 counts in the three detectors.
For these stars, we computed the errors in $T_{\rm av}$ as follows: 
We determined the 68\% confidence contour  on the $\beta-T_0$ plane for 
these two ``parameters of interest'', i.e. the loci for which a fit
can be achieved whose $\chi^2$ is larger by $\Delta \chi^2 = 2.3$
(1 $\sigma$) than the $\chi^2$ of the best fit. We then found the minimum 
and the maximum $T_{\rm av}$ for this subset of models, and thus defined the
error range for $T_{\rm av}$.  
Using these errors, we computed the weighted mean of $\log T_{\rm av}$.
For CTTS, we neglected DD Tau AB, which shows extraordinarily
high temperatures in two different observations (two bullets at the
hottest temperature in Fig.~\ref{lx_tav}). For CTTS, we find 
$\langle \log T_{\rm av} \rangle = 6.97 \pm 0.06$ ($6.98 \pm 0.06$ 
if DD Tau AB is also considered) while for WTTS we find $\langle \log 
T_{\rm av} \rangle = 6.81 \pm 0.05$. We therefore conclude that
the CTTS and WTTS with $L_{\rm X} < 3 \times 10^{29}$~erg s$^{-1}$ are different
at a 3$\sigma$ level, fully supporting the significant differences in the regression
fits that are based on the entire $L_{\rm X}$ range.

In conclusion, we find the CTTS X-ray sources to be hotter than WTTS
at a confidence level of $\gtrsim 98$\%, and this
result is partly due to the presence of a $L_{\rm X}-T_{\rm av}$
relation for WTTS but its absence in CTTS. Further, the WTTS relation coincides
with relations valid for main-sequence stars of different spectral types,
including saturated and non-saturated
stars at different evolutionary stages.

\section{Discussion}\label{discussion}

\subsection{Summary of trends} 

We now discuss the trends and correlations described in the previous section
and will also put them into a context with previous reports, in particular
from the COUP project. 

The most significant correlations that we reported above are those between
stellar mass and $L_{\rm X}$ (slope $\approx$ 1.7), between stellar bolometric 
luminosity $L_*$ and $L_{\rm X}$ (slope $\approx 1$), and between $L_{\rm X}$ and average electron 
temperature $T_{\rm av}$ (slope 0.15--0.23), the latter applying only to WTTS.

Further, we have found that $L_{\rm X}$ and $L_{\rm X}/L_*$ are both lower, on average,
for CTTS than for WTTS, each by a factor of $\approx 2$, compatible with 
the finding that the distributions of $L_*$ are similar for the two samples. 
In contrast, $T_{\rm av}$ is, on average, higher by a factor of $\approx 1.7$ for 
CTTS than for WTTS.

Finally, we have studied possible correlations between $L_{\rm X}, T_{\rm av}$, or
$L_{\rm X}/L_*$  and the accretion rate but found at best unconvincing correlations. The same is true for a trend
between $L_{\rm X}$ and age.

\subsection{Comparison with previous studies} 

{\bf The $L_{\rm X}$ -- mass correlation.} This relation has been reported prior to
{\it XMM-Newton} and {\it Chandra} studies of star-forming regions, but with largely
varying regressions. \citet{feigelson93} found a slope of $3.6\pm 0.6$ for
a sample of low-mass stars in the Chamaeleon I dark cloud based on ROSAT observations.
There may be problems with more numerous upper limits at the low-mass (and low-luminosity)
end of the distribution in this study, as noticed by \citet{preibisch05}. On the other
hand, the COUP sample reveals a very similar correlation to ours, with a slope only marginally
smaller ($1.44\pm 0.10$) than for XEST ($1.69\pm 0.11$). The TMC sample thus essentially
confirms the COUP results, and the residual difference might be due to a somewhat
different distribution of stars in the HRD, perhaps indicating a different age 
distribution, as suggested from our discussion of this relation below.

A clear difference between the two studies is seen in the scatter around the regression
curves. While \citet{preibisch05} report a standard deviation of 0.65~dex (factor of 4.5)
around the best-fit line, we find for the XEST sample values between 0.38 (factor of 2.4, for
WTTS) and 0.45 (factor of 2.8, for CTTS). We do not have a clear explanation for the 
smaller scatter in XEST, but note that i) similar findings apply to other correlations
discussed below, and ii)  the scatter found in the XEST results is close to the intrinsic
uncertainties of {\it any} $L_{\rm X}$ measurement of magnetically active stars as these commonly
vary by such factors on various time scales; \citet{preibisch05} give a factor of 2 variation
on long (yearly) time scales for the Orion sample. We are thus confident that the quality
of our mass  -- $ L_{\rm X}$ correlation corresponds to the minimum scatter that must be expected 
from snapshot observations of magnetically active stars.

{\bf The $L_{\rm X} - L_*$ correlation.} The linear correlation between these variables
expresses the classical result of {\it X-ray saturation} that has been found empirically
for main-sequence stars \citep{vilhu83}. A similar law applies to very active main-sequence
and subgiant stars (see review by \citealt{guedel04b} and references therein), and certainly
also to pre-main sequence stars at various evolutionary stages \citep{flaccomio03b}. 
Again, the COUP study is in complete agreement with our results, its regression
slope being $1.04 \pm 0.06$. There is, however, a significant difference between
the XEST and the COUP studies once CTTS and WTTS are treated separately. \citet{preibisch05}
find a well-defined linear correlation for WTTS (standard deviation around best-fit regression
of 0.52~dex), while for CTTS the scatter dominates (standard deviation = 0.72 dex) and the 
relation is significantly flatter. The CTTS data points span a range of 3 orders of magnitude at 
a given $L_*$. In XEST,  the standard deviation of the scattered points
is only $\approx 0.4$ dex for CTTS, WTTS, and the entire sample, with a range of $L_{\rm X}$ values at a 
given $L_*$ of about 1.5~dex. We are again not in a situation to explain the much tighter
correlations for the XEST survey, but note that our spectral-fit methodology may suppress
numerical uncertainty introduced by photoelectric absorption that suppresses evidence of plasma
components at lower temperatures. Because  we used an emission-measure distribution with
a prescribed low-temperature shape as usually found in magnetically active stars, the presence
of the coolest components is interpreted based on the presence of well-detected hotter plasma.
An error analysis for $L_{\rm X}$ based in particular on $N_{\rm H}$ shows that for 76\% of
the sources in XEST, the intrinsic error range in $L_{\rm X}$ due to $N_{\rm H}$
is smaller than a factor of 3 (0.5~dex, in fact mostly much smaller), and the largest 
errors are obtained for faint sources subject to $N_{\rm H} > (2-3)\times 10^{21}$~cm$^{-3}$
\citep{guedel06a}.
We note that \citet{preibisch05} used an X-ray luminosity averaged over the
10 days of exposure by {\it Chandra}, while in the XEST sample we have neglected time
intervals containing obvious flares. However, \citet{preibisch05} found that the average
$L_{\rm X}$ and the quiescent (``characteristic'') $L_{\rm X}$ differ by a median factor
of 0.78, which would not be sufficient to explain the large scatter found in their
$L_{\rm X} - L_*$ correlation for CTTS.

{\bf Comparison with main-sequence stars.}  \citet{preibisch05} also compare their
$L_{\rm X} - L_*$ correlation with field main-sequence stars, and find a much shallower 
slope for the latter, but also a very large scatter. They similarly compare the 
$L_{\rm X} - $ mass relations with field stars, but there, they find a similar slope 
in the regression. We will not perform this comparison  here, for the following reason.
Field stars are found at various evolutionary stages, and as a consequence of  stellar
spin-down with age, the X-ray activity is subject to an evolutionary decay. Solar analogs 
decrease in $L_{\rm X}$ by three orders of magnitude from the zero-age main sequence to
the end of the main-sequence life \citep{guedel97, telleschi05}. Further, the evolutionary speed is 
different for G stars and low-mass M dwarfs, the latter remaining at relatively high activity
levels for a longer time (see Figs. 40 and 41 in \citealt{guedel04b}). Much of the scatter in 
$L_{\rm X}$ for a given mass or a given $L_*$ is thus due to mass-dependent evolutionary decay, 
and any trend in $L_{\rm X}$ vs. $L_*$ depends on the stellar age distribution. In contrast,
both in active main-sequence stars and TTS, no evolutionary effects are expected for 
the $L_{\rm X} - L_*$ relation if the stars are in a saturated regime, and
therefore $L_{\rm X} \propto L_*$. 

For the $L_{\rm X}$ -- mass relation,
the scatter in $L_{\rm X}$ for a given mass is only about one order of magnitude;
this scatter is indirectly due to the scatter in $L_*$ in the sample, due to
different ages of stars of similar mass that contract vertically along 
the Hayashi track, provided that the X-ray emission remains in a saturated state 
(see HRD in Fig.~11 in \citealt{guedel06a}). The scatter
in $L_{\rm X}$ due to evolution on the main sequence is much larger (3 orders of magnitude)
and is due to intrinsic decay of the dynamo due to stellar spin down when the X-ray emission
is no longer in a saturated state.

{\bf The $L_{\rm X} - T_{\rm av}$ correlation.} A dependence between coronal electron 
temperature and emission measure or $L_{\rm X}$ (or normalized quantities such as
the specific emission measure or surface X-ray flux) has first been noted by \citet{vaiana83}
and \citet{schrijver84}. Quantitatively, for solar analogs, $L_{\rm X} \propto T^{4.5\pm 0.3}$
(\citealt{guedel97}, see \citealt{guedel04b} for a review). For the pre-main sequence 
sample in the COUP survey, \citet{preibisch05} report a steep increase of the X-ray surface flux
with the hotter temperature of their 2-component spectra, namely $F_{\rm X} \propto
T_2^6$. On the other hand, they find a relatively constant lower temperature, namely $T_1 \approx
10$~MK. In our study, we apply a more physically appropriate continuous emission measure
distribution that does not distinguish between two isothermal components but that 
shows two power-law slopes on either side of the peak. The distribution of the logarithmically
averaged temperatures (Fig.~\ref{tav_kaplan}) does not show a preferred value but a smooth 
distribution in the range 4--30~MK around a mean of 7.6~MK for WTTS and 12.6~MK for CTTS 
(Sect.~\ref{sect_temp}). Our regression curve for the  $L_{\rm X} - T_{\rm av}$ indicates 
$L_{\rm X} \propto T_{\rm av}^{4.3-6.7}$ and $F_{\rm X} \propto T_{\rm av}^{3.8-5.6}$,  
compatible with the solar-analog relation as well as with the COUP relation for $T_2$. We note,
however, that we find this trend only for WTTS (no separate analysis was provided for COUP). 

{\bf $\log (L_{\rm X}/L_*)$ distributions.} Our distributions show the fractional X-ray luminosity
of CTTS to be suppressed by a factor of $\approx 2$ compared to WTTS, with a mean 
$\log(L_{\rm X}/L_*) \approx -3.39\pm 0.06$ and $-3.73\pm 0.05$ for WTTS and CTTS, respectively. 
These values agree excellently with those of COUP: $\log(L_{\rm X}/L_*) \approx -3.31 $ 
and  $-3.73$ for WTTS and CTTS, respectively \citep{preibisch05}. XEST contrasts with
COUP in that CTTS are less X-ray efficient for all considered mass bins, whereas 
\citet{preibisch05} found no difference in the ranges 0.1-0.2 $M_{\odot}$ and
0.5-1 $M_{\odot}$.

\subsection{The $L_{\rm X}$ -- mass relation}

\begin{figure}
\centering
\includegraphics[angle=-0, width=0.49\textwidth]{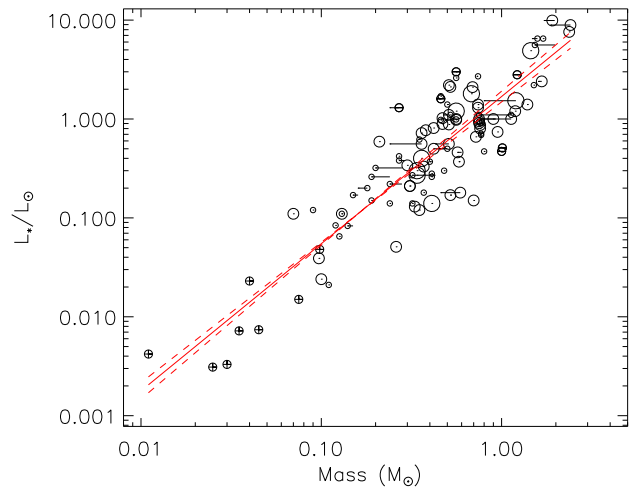}
\caption{Stellar bolometric luminosity as a function of mass. The straight line 
     is a Y/X OLS regression fit; dashed lines illustrate the 1-sigma errors in the slope.}
      \label{m_lbol}
\end{figure}

Among the clearest correlations we have identified is the $L_{\rm X}$-mass relation
that closely corresponds to the finding in the COUP study. We now test the following:
If we assume that TTS are in a saturated state (i.e. $L_{\rm X} \propto L_*$, Fig.~\ref{lx_lbol})
and a relation between $L_*$ and stellar mass exists (related to the age distribution
of the stars and details of the evolution along the pre-main-sequence tracks), then the relation 
between $L_{\rm X}$ and mass could simply be a consequence of these two relations.
Main-sequence stars follow the well-known mass-bolometric luminosity relation,
which for stars in the mass range of 0.1--1.5$M_{\odot}$ reads $L_* \propto M^{3.0}$
(from a regression analysis using the \citealt{siess00} ZAMS data). For pre-main sequence
stars, a mass-bolometric luminosity relation is not obvious; during the contraction phase, 
a star of a {\it given mass} decreases its $L_*$ by up to 2 orders of
magnitude. However, if most stars in a sample show similar ages, then an approximate
mass-$L_*$ relation may apply to the respective isochrone. 
Fig.~\ref{m_lbol} illustrates the {\it measured} relation between $L_*$ and mass. The relation
is rather tight, with a correlation coefficient of 0.90 for 113 sources.
The Y/X OLS regression gives $\log L_*/L_{\odot} = (1.49 \pm 0.07) \log M
+ (0.23 \pm 0.04)$, with a standard deviation of 0.31. 
This relation can be compared with the theoretical prediction for
an average isochrone appropriate for the XEST sample. We found
that the logarithmically averaged age of our targets is 2.4 Myr.  
Adopting the Siess et al. (2000) isochrones, we find, from a linear regression
fit, a dependence $L_* \propto M^{1.24}$, i.e., similar to the observed
dependence and thus supporting our interpretation. We note that the
XEST sample is of course not located on an isochrone (see Fig. 11 in \citealt{guedel06a}), and that
other evolutionary calculations may lead to somewhat different
slopes of the isochrones.

\begin{figure}
\centering
\includegraphics[angle=-0, width=0.49\textwidth]{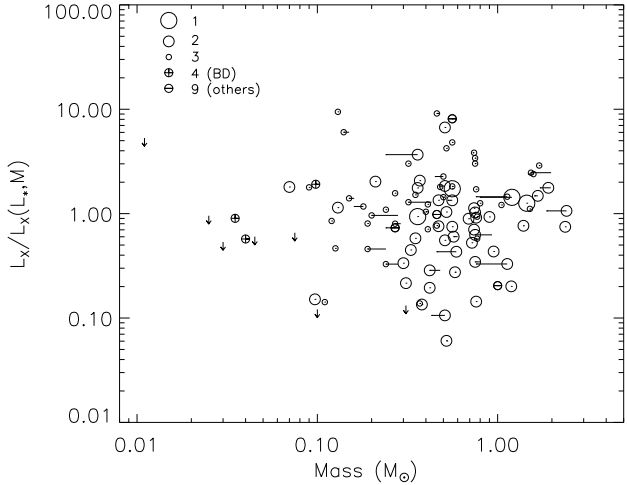}
\caption{$L_{\rm X}$ as a function of mass after renormalizing $L_{\rm X}$ with the expected 
         $L_{\rm X}$ based on mass-bolometric luminosity relation and the saturation law 
         (see text for details).  Different symbols mark different types of stars.}
      \label{mlx_sub}
\end{figure}

Adopting  $\log (L_{\rm X}/L_*) = -3.5$ for our entire TTS sample (see Fig.~\ref{lx_lbol}), we infer 
a relation $\log L_{\rm X} = 1.49 \log M +30.31$,
similar to the correlation found in Sect.~\ref{sect_mass}.
In Fig.~\ref{mlx_sub} we plot $L_{\rm X}$ as a function of mass 
after normalizing the observed $L_{\rm X}$ with $L_{\rm X}$ predicted from
the above formula. The correlation found
in Fig.~\ref{mass_lx} now disappears completely. The scatter
in Fig.~\ref{mlx_sub} is due to the scatter in Fig.~\ref{lx_lbol}
and in Fig.~\ref{m_lbol}, i.e., intrinsic scatter not due to $L_*$ or
$M$, but for example due to the evolutionary decrease of $L_*$ and $L_{\rm X}$ along 
the Hayashi track for a given mass. 

We therefore conclude that the $L_{\rm X}$-mass relation is not an intrinsic 
relation but a consequence of an approximate mass-luminosity relation for stars 
with similar ages, combined with a saturation law.

Comparing the stars in the HRD of the TMC (Fig. 11 in  \citealt{guedel06a})
 with the HRD of the ONC sample (Fig. 1 in \citealt{preibisch05}), we note that
 the ONC sample is somewhat younger, although it is not tightly arranged 
 along one isochrone, but tends to show a somewhat flatter slope than the TMC sample. 
 The slightly shallower $L_{\rm X}$-mass relation in \citet{preibisch05} may therefore
 be a straightforward consequence of the younger average age of the ONC sample.

\subsection{Origin of the X-ray emission: CTTS vs WTTS}

The question that we address in this section is
on the origin of the X-ray emission. How do CTTS and WTTS 
differ, and what may be the causes? Where and how are the 
X-rays formed?  

The relevant relations we have identified in this paper are the following:
i) CTTS show, on average, a smaller X-ray luminosity in the EPIC band;
ii) CTTS also reveal a significantly lower fractional X-ray luminosity,
$L_{\rm X}/L_*$, than WTTS. iii) The average electron temperature
in WTTS correlates with the total $L_{\rm X}$ for WTTS, but this
is not the case for CTTS; the CTTS temperatures are on average significantly
higher than those of WTTS. 

\subsubsection{Evidence for coronal emission}

The bulk of the X-ray emission described in this paper and in the
COUP survey is consistent with an origin in a magnetic corona.
The overall temperatures measured in TTS and the X-ray luminosities are
similar to values also found in extremely active main-sequence 
and subgiant stars (e.g., review by \citealt{guedel04b}). Also, frequent
flaring in TTS \citep{wolk05, stelzer06, franciosini06} clearly
points to a coronal (or magnetospheric) origin of the X-ray emission.
Although many of the T Tauri stars in the XEST sample are thought to be 
fully convective and hence unable to support a solar-like $\alpha$-$\Omega$ 
dynamo, fully-convective main-sequence stars do show magnetic activity, and 
dynamo mechanisms have been proposed that may operate in such stars (e.g., 
\citealt{dobler06} and references therein; 
\citealt{kueker99}). A discussion of what XEST can tell us about the 
dynamos acting in T Tauri stars in Taurus-Auriga is given by \citet{briggs06}. 

We have found in the previous section that the  $L_{\rm X}$ - mass relation is due to
saturation and a mass-bolometric luminosity relation for TMC stars.
The most fundamental relation is indeed the one between $L_{\rm X}$
and $L_*$, which indicates that the PMS stars in Taurus
are saturated at $\log (L_{\rm X}/L_*) \approx -3.5$, in analogy to
main-sequence stars. This result suggests that the
bulk of X-ray emission in PMS stars by CCD detectors used in XEST and in COUP
arises from a magnetic corona, and we therefore suggest that similar processes as 
in MS stars should be responsible for the dominant energy output from the hot plasma 
(we will rediscuss emission from the softest X-ray emitting plasmas below).

\subsubsection{Is a lower $L_{\rm X}$ intrinsic to CTTS coronae?}


We have tested (Sect.~\ref{sec_xlf}) whether the lower $L_{\rm X}$ seen in CTTS 
compared to WTTS could be a consequence of generally lower $L_*$ of CTTS, provided
that all stars are in a similar saturation regime. This can be explicitly rejected
because we found the $L_*$ distributions for the two samples to be similar, with
a trend that it is rather the CTTS sample that is slightly more luminous.
Also, the $L_{\rm X}/L_*$ distributions are explicitly different, offset by
about the same factor as the $L_{\rm X}$ distributions themselves. 

In previous works the lower X-ray activity of CTTS compared to WTTS in 
Taurus-Auriga has been attributed to the slower rotation of CTTS and an 
anticorrelation of activity and rotation period as exhibited by active 
solar-like stars (e.g., \citealt{neuhaeuser95}). However, the lower X-ray 
activity of CTTS has been observed in other star-forming regions where an 
anticorrelation of activity and rotation period is clearly not seen (e.g. 
\citealt{preibisch05}). \citet{briggs06} demonstrate that an apparent 
activity-rotation relation in Taurus-Auriga naturally results from the 
dependences of activity on mass and accretion status reported here and in 
other star-forming regions because the fast rotators in Taurus-Auriga
are mainly higher-mass and non-accreting while the slow rotators are mainly 
lower-mass and accreting. There 
is no convincing evidence for an anticorrelation of X-ray activity 
and rotation period in T Tauri stars, and therefore no evidence that the 
lower activity of CTTS is due to their slower rotation.
 
Indeed, even if a solar-like dynamo operates in T Tauri stars, their long 
convective turnover timescales lead to the expectation that all stars with 
measured rotation periods should have saturated (or supersaturated) 
emission (e.g., \citealt{preibisch05}) and show no anticorrelation of 
activity and rotation period.

Different internal structure in WTTS and CTTS (unless induced by accretion
processes) is not a likely explanation either. As was recognized in early 
infrared and optical surveys of the TMC stellar population, WTTS and CTTS 
occupy the same region in the HRD \citep{kenyon95}, with no evolutionary 
separation, indicating that the transition from CTTS to WTTS occurs at very 
different ages for different stars. 

This suggests that the lower $L_{\rm X}$ of CTTS is an intrinsic
property of a corona that is heated in the presence of an accretion
disk and active accretion onto the star. We now briefly consider the
influence of accretion  on coronal heating.

\subsubsection{The role of active accretion}

Accretion has variously been suggested to enhance or to suppress
plasma heating. First, accretion hot spots may heat plasma to temperatures in
excess of one million K as gas shocks near the surface in nearly free fall
(e.g., \citealt{calvet98}). High-resolution X-ray spectroscopy has
provided some indirect evidence that such accretion-induced X-rays might constitute
an important part of the measurable spectra. \citet{kastner02} have
interpreted exceptionally high-densities and very cool ($T \approx 3$~MK)
X-ray emitting material in the CTTS TW Hya as being the result of 
accretion shocks. This idea was further elaborated by \citet{stelzer04} 
who also suggested that anomalously high Ne and N abundance in this
X-ray source indicates that refractory elements such as Fe condense
onto dust grains in the disks, and that this material is eventually not accreted. A 
small number of additional X-ray spectra have been studied, but the situation is 
complex and contradictory: \citet{schmitt05} and \citet{robrade06} report
intermediately high densities in BP Tau ($n_{\rm e} \approx 3\times 10^{11}$~cm$^{-3}$), 
and \citet{guenther06}  find similarly high densities in V4046 Sgr.
On the other hand, very low densities and no strong abundance anomalies have been found in T Tau 
\citep{guedel06c} and the accreting Herbig star AB Aur \citep{telleschi06b}. 
However, \citet{telleschi06a} found evidence that CTTS in general maintain
an excess of cool (1-4~MK) plasma compared to WTTS. This expresses itself
in O\,{\sc vii} line fluxes that are similar to the flux in the
 O\,{\sc viii}~Ly$\alpha$ line, a condition very different  from WTTS where
the O\,{\sc vii} lines are very faint (similar to magnetically active ZAMS stars).

This soft excess seems - as far as the still small statistical sample of stars
suggests - to be related to accretion in CTTS. But where
are the X-rays produced? For T Tau and AB Aur, a production in accretion 
shocks is unlikely \citep{guedel06c, telleschi06b} while accretion shocks 
may be responsible for the softest emission in TW Hya, BP Tau, and V4046 Sgr 
\citep{kastner02, stelzer04, schmitt05, guenther06}. 

\subsubsection{Coronal modification by accretion streams?}

An alternative possibility is that accretion influences the magnetic 
field structure of the stars, or that the accreting material
is changing the heating behavior in the coronal magnetic fields. 
\citet{preibisch05} proposed that the mass-loaded loops of accreting 
stars are denser than the loops of WTTS, so that when a magnetic
reconnection event occurs, the plasma would be heated to 
much lower temperatures outside the X-ray detection limit. This could 
account for the deficiency of X-ray luminosity in CTTS compared to
WTTS. A similar scenario has been proposed by \citet{guedel06c} to specifically
explain  the soft excess observed in T Tau. In this scenario, the 
magnetospheric geometry is influenced by the accretion stream. A fraction 
of the cool accreting material enters the coronal active regions
and cools the magnetic loops there for three reasons: i) the accretion flow
may reorganize magnetic fields, stretching them out and making them less 
susceptible to magnetic reconnection; ii) the accreting material is
cold, lowering the resultant temperature on the loops when mixing with
the hot plasma; and iii) the accreting material itself adds to
the coronal density, inducing larger radiative losses and more rapid 
cooling. These resulting soft X-rays from cool material can be detected  
by the RGS instruments that are sensitive at low energies and provide the
spectral resolution to record flux from lines formed at temperatures 
below 3~MK, such as O\,{\sc vii} and N\,{\sc vi}, but the same 
emission will not be separately identified by the EPIC detectors that
are relatively insensitive at the relevant temperature, and provide 
only very low energy resolution. \citet{guedel06c} estimated for T Tau
that only of order 1\% of the accreting material would be needed to 
penetrate active regions on the star and be heated to 2~MK.

If the total coronal energy release rate (averaged over time scales 
longer than energy release events such as flares) is determined by
the stellar dynamo that forms a magnetic corona (and by convective
properties near the stellar surface), then we would expect similar
coronal radiative losses for CTTS and WTTS. Could it be that the radiative output from
the corona is indeed equivalent but that part of the coronal 
emission has shifted to the softest part of the spectrum, remaining
undetected in EPIC CCD spectroscopy while detected as a soft excess
by RGS? At least in the case of T Tau, the 0.3-10~keV
X-ray luminosity has been severely underestimated by EPIC CCD analysis alone,
as these instruments missed the softest component, also subject to
considerable photoelectric absorption, and suggested $L_{\rm X}$ to
be only $\approx 60$\% of $L_{\rm X}$ determined from the combined 
RGS+EPIC spectra \citep{guedel06c}.

We systematically studied our RGS spectra (from \citealt{telleschi06a}) 
to find out whether the soft excess in our CTTS sample provides
the ``missing luminosity''. We have therefore compared the unabsorbed
0.1-0.5~keV  X-ray luminosity  from the EPIC spectral fits \citep{guedel06a}
with $L_{\rm X}$ in the same range from the combined EPIC+RGS
fits \citep{telleschi06a}. The comparison is useful only for targets that
do not suffer from strong absorption; this is the case for the CTTS BP Tau and DN Tau and the
WTTS HD~283572, V773 Tau, and V410~Tau. We found no systematic difference to
explain the factor of 2 underluminosity of CTTS. 
It appears that the EPICs record  the soft emission from the coolest plasma
sufficiently well to register similar $L_{\rm X}$ as the RGS detectors, but the temperature 
discrimination is clearly inferior to high-resolution spectroscopy. Also, the softest
range is still dominated by continuum emission from hotter plasma, and the soft excess
in these stars provides relatively little spectral flux. 

We have found, on the other hand, that CTTS show, on average, higher
electron temperatures (averaged over the components detected by the
EPIC cameras) than WTTS. This could be an effect of depletion of
the intermediate and cooler temperature ranges by the accretion process,
as suggested above, thus moving the average temperature of the
{\it detected} coronal components to higher temperatures. We have therefore
tested whether the harder portion of the EPIC spectra which is radiated
by the hottest coronal plasma components also shows a statistical difference
between CTTS and WTTS. We chose the 1.5--10~keV range for this test.
We found, however, the same discrepancy between the two stellar groups, suggesting that the emission
measures of the hottest plasma components themselves are also suppressed in
CTTS compared to WTTS.

We extended our  comparison
to the XEST results obtained from EPIC only, but again found CTTS to be underluminous 
by similar factors in {\it different} X-ray energy bands. To explain the deficit of X-ray emission 
in CTTS, it could thus be that the accretion  process is cooling active-region plasma to 
an extent that it is also no longer detected in the RGS band. 

\subsubsection{Coronal heating in WTTS and CTTS due to flares?}

This brings us to the correlation between average coronal temperature
and $L_{\rm X}$ which is present in WTTS but absent in CTTS. WTTS
show a trend in which $T_{\rm av}$ increases with $L_{\rm X}$, and this
trend is the same as previously found for main-sequence solar analogs
\citep{guedel97, telleschi05}. We note that this relation remains valid
even for stars with different X-ray saturation limits and different $L_*$
(see Fig.~\ref{lx_tav}). M dwarfs at the saturation limit reveal lower coronal
temperatures than K or G dwarfs at their saturation limit. The cause for
this relation is not clear. \citet{guedel04b} pointed out that
the slope of the regression function (using emission measure instead
of $L_{\rm X}$) is the same as the slope of the regression between
peak temperature and peak emission measure in stellar flares. \citet{guedel04b}
hypothesized that coronal emission is formed by a superposition of 
continuously occurring ``stochastic flares'', with the consequence that
larger, hotter flares that occur more frequently in more active  stars 
not only produce the dominant portion of the observed emission measure, 
but also heat the observed plasma to higher temperatures than in lower-activity stars.
The larger rate of large flares would be a consequence of denser packing of magnetic fields
in more active stars, inducing more frequent explosive magnetic reconnection,
including larger flares than in low-activity stars \citep{guedel97}. A similar trend for WTTS as for
main-sequence stars is therefore perhaps not surprising: X-ray production in both
types of stars is thought to be entirely based on the magnetic field production by the
internal dynamo. This analogy fully supports solar-like coronal processes in WTTS.

The $L_{\rm X} -T_{\rm av}$ correlation is absent in CTTS. The distinguishing property of 
CTTS is active accretion, which thus is most likely the determining factor for the
predominantly hot coronal plasma. Temperatures like those determined as $T_{\rm av}$ in 
the XEST survey cannot be produced in accretion shocks, again pointing to a predominantly
coronal origin of the hot, dominant plasma component. If the flare-heating concept
has merit in CTTS as well, then it seems that flares in CTTS are predominantly hot, even
if $L_{\rm X}$ is low. We can only speculate about the origin of this feature. A possibility 
are star-disk magnetic fields, but it is unclear why flares occuring in
such loop systems should be hotter.

\section{Summary and conclusions}\label{conclusions}

We have studied X-ray parameters of a large sample of X-ray spectra of CTTS and WTTS
in the Taurus Molecular Cloud. Our principal interest has been in a characterization of
X-rays in the two types of stars, in finding correlations between X-ray parameters and
fundamental stellar properties and among themselves, and most importantly in 
comparing our findings between accretors and non-accretors. This study has been
motivated by numerous previous reports on correlations and differences between CTTS
and WTTS in nearby star-forming regions, and in particular by the COUP study of 
the Orion Nebula Cluster. The XEST project has provided the deepest and, for the surveyed
area, most complete X-ray sample in the Taurus region to date. We have used
a CTTS and a WTTS sample of comparable size.

We have correlated $L_{\rm X}$ and the average coronal temperature, $T_{\rm av}$, with
various stellar parameters, and conclude the following from our study:

\begin{itemize}

\item The X-ray luminosity is well correlated with the stellar mass, with a dependence $L_{\rm X} \propto M^{1.7}$, 
      similar to what has been shown in COUP and previous TTS studies, but we find that
      this correlation is only an expression of saturation and a mass-(bolometric) luminosity
      relation for our pre-main sequence sample. As long as the stellar sample is saturated,
      $L_{\rm X}$ is a function of $L_*$, and the latter is correlated with stellar mass for
      a given isochrone. From stellar evolution calculations (e.g., \citealt{siess00}),
      the functionality between mass and $L_*$ can be derived. This is fully analogous
      to main-sequence stars where approximately $L_* \propto M^{3}$ holds. 
      For a typical isochrone of TMC stars with ages of 2--3 Myr, the exponent is smaller
      as can be seen on a pre-main sequence HRD (Fig.~11 in \citealt{guedel06a}).
      For our sample, $L_* \propto M^{1.5}$.
      
\item A saturation relation holds for both CTTS and WTTS,  although $L_{\rm X}/L_*$ is, on
      average, smaller by a factor of 2 for CTTS compared to WTTS.
      
\item We find that the distributions of $L_*$ are similar for CTTS and WTTS. As a consequence,
      we find a significant difference in the X-ray luminosity functions for CTTS and WTTS,
      the former being fainter by about a factor of two. The suppressed X-ray production
      in CTTS is thus intrinsic to the source and not due to selection bias.
    
\item We emphasize that the lower X-ray production in CTTS {\it refers to the range of plasma
      temperatures accessible by CCD cameras such as those used here and in COUP}. It is possible
      that some of the energy release is shifted to lower temperatures outside
      the range easily accessible to CCD detectors. Those soft regions of the X-ray spectrum are
      also subject to increased photoelectric absorption, which makes detection of cool
      plasma more difficult. CTTS are indeed more absorbed than WTTS, namely by 
      a factor of $\approx 2.5$.
      
\item We investigated whether all X-ray spectral ranges show X-ray suppression in CTTS.
      The hardest  portion  (1.5--10~keV) shows the X-ray deficiency
      in CTTS vs. WTTS independently, even though CTTS reveal higher average
      temperatures. We hypothesized that a fraction of the emission measure has been
      cooled to poorly detectable or undetectable temperatures in CTTS.
      CTTS indeed show a {\it soft excess} in their high-resolution 
      X-ray spectra, characterized by unusually strong O\,{\sc vii} lines from cool plasma
      \citep{telleschi06a}. These lines cannot be resolved by EPIC. We therefore
      checked whether RGS spectroscopy of little absorbed CTTS stars in \citet{telleschi06a} 
      (DN Tau, BP Tau) indicates a relative increase of $L_{\rm X}$  in the soft 0.1--0.5~keV band
      relative to WTTS, but found no significant effect. The soft flux may have been sufficiently well
      detected by EPIC in these low-absorption stars (but with little temperature discrimination). 
      Also, the softest range is still dominated by continuum emission from hotter plasma, 
      and the soft excess in these stars provides relatively little spectral flux. The X-ray 
      deficiency in CTTS thus remains.\\
      The situation is clearly different in T Tau \citep{guedel06c}: in this much more strongly
      absorbed source, a very large amount of very cool X-ray emitting plasma was detected based 
      exclusively on anomalously strong O\,{\sc vii} 
      line emission in the grating spectrum but went unnoticed 
      in CCD spectroscopy. The analysis of the latter spectrum alone led to an underestimate 
      of the 0.3--10~keV luminosity by 40\%.   \\
      In conclusion, it seems that the entire X-ray range accessible to CCD spectroscopy reveals
      suppressed X-ray emission compared to WTTS, although additional components may be
      present at cool temperatures that may be missed by the CCD spectra, especially if 
      $N_{\rm H}$ is sufficiently high.
      
\item A possible cause for the suppression of X-ray emission in CTTS may be the 
      accretion streams themselves. If only a small portion of the accreting matter
      penetrates into hot coronal magnetic structures, the plasma may cool as more matter
      needs to be heated and as the increase in density increases the cooling efficiency.
      This may lead to a soft excess \citep{guedel06c}, or to a cooling of 
      plasma to temperatures outside the X-ray regime \citep{preibisch05}, so that
      a significant deficiency of X-ray emission may be measured in the spectral
      range that is accessible to CCD cameras, and that is not subject to significant 
      photoelectric absorption. {\it The soft excess and the hot-plasma deficiency
      seem to be related to the presence of accretion.}

\item X-ray production in shocks at the base of accretion streams has been suggested
      previously from high-resolution spectra of CTTS. The shocked plasma would add soft emission
      to the spectra as well, but again, CCD spectroscopy may miss this emission, or
      the latter may be subject to absorption. Our CCD survey does not provide the
      appropriate means to test X-ray production in accretion shocks in CTTS, and
      can therefore also not exclude such mechanisms. High-resolution  grating spectroscopy is
      required.

\end{itemize}

\begin{acknowledgements}

We are grateful to Brian E. Wood for providing emission measure distributions of 
intermediately active K stars (used in Fig.~\ref{lx_tav}).
We acknowledge helpful comments by the referee, J. Kastner.
We thank the International Space Science Institute (ISSI) in Bern,
Switzerland, for logistic and financial support during several workshops on the TMC 
campaign. This research is based on observations obtained with {\it XMM-Newton}, an 
ESA science mission with instruments and contributions directly funded by ESA Member 
States and the USA (NASA).
X-ray astronomy research at PSI has been supported by the Swiss National Science
Foundation (grant 20-66875.01 and 20-109255/1).
 M.~A. acknowledges support by National Aeronautics and Space Administration 
 (NASA) grant  NNG05GF92G. In addition, he acknowledges support from a Swiss 
 National Science Foundation Professorship (PP002--110504).

\end{acknowledgements}
%
%

\end{document}